\documentclass[12pt]{article}

\ifx\pdfoutput\undefined
\usepackage[dvips,bookmarks]{hyperref}
\else
\usepackage{hyperref}
\fi
\hypersetup{colorlinks=false,bookmarksopen,bookmarksnumbered,citecolor=blue,
   pdfstartview=FitH}

\usepackage{latexsym}
\usepackage{amssymb,amsfonts,amsmath}
\usepackage{graphicx} 
\usepackage{indentfirst}
\usepackage{bbm}
\usepackage{amssymb}
\usepackage{verbatim}
\usepackage{amsmath, amsthm,amssymb}
\usepackage{mathrsfs}
\usepackage{hyperref}
\usepackage{amsfonts}
\usepackage{dsfont}
\usepackage{slashed, tensor}
\usepackage{booktabs}
\usepackage{graphicx}
\usepackage{mathrsfs}
\usepackage[colorinlistoftodos,prependcaption,textsize=small]{todonotes}
\parskip=\medskipamount

\newcommand {\cA}{{\cal A}}

\newcommand {\cG}{{\cal G}}
\newcommand {\cH}{{\cal H}}

\newcommand {\cK}{{\cal K}}
\newcommand {\cL}{{\cal L}}
\newcommand {\cM}{{\cal M}}
\newcommand {\cN}{{\cal N}}
\newcommand {\cO}{{\cal O}}

\newcommand {\cQ}{{\cal Q}}

\newcommand {\cS}{{\cal S}}
\newcommand {\cT}{{\cal T}}

\def\a{\alpha}
\def\b{\beta}
\def\c{\chi}
\def\d{\delta}
\def\e{\epsilon}
\def\f{\phi}
\def\g{\gamma}

\def\k{\kappa}
\def\l{\lambda}

\def\o{\omega}

\def\q{\theta}
\def\r{\rho}
\def\s{\sigma}
\def\t{\tau}

\def\J{\Psi}
\def\L{\Lambda}

\def\S{\Sigma}
\def\U{\Upsilon}
\def\X{\Xi}
\def\tr{{\rm tr}}

\def\ri{{\rm i}}
\def\re{{\rm e}}

\newcommand{\ad}{{\dot{\alpha}}}                      
\newcommand{\bd}{{\dot{\beta}}}                            
\newcommand{\ve}{\varepsilon}

\newcommand{\ab}{{\a\b}}

\newcommand{\pa}{\partial}            
\newcommand{\hf}{\frac12}
\newcommand{\vf}{\varphi}
\newcommand{\be}{\begin{equation}}
\newcommand{\ee}{\end{equation}}
\newcommand{\bea}{\begin{eqnarray}}
\newcommand{\eea}{\end{eqnarray}}
\newcommand{\non}{\nonumber}
\newcommand{\ba}{\begin{array}}
\newcommand{\ea}{\end{array}}

\newcommand{\bm}[1]{\mbox{\boldmath$#1$}}

\def\double #1{#1{\hbox{\kern-2pt $#1$}}}

\newcommand{\hm}{{m}}
\newcommand{\hn}{{n}}

\newcommand{\bbD}{{\mathbb {D}}}

\newcommand{\bsubeq}{\begin{subequations}}
\newcommand{\esubeq}{\end{subequations}}

\newcommand{\ul}{\underline}
\newcommand{\eps}{{\ve}}

\newcommand{\rd}{\mathrm d}
\numberwithin{equation}{section}
\newcommand{\RM}{R(M)}
\newcommand{\RD}{R(\mathbb D)}

\newcommand{\RJ}{R(J)}
\newcommand{\RS}{R(S)}
\newcommand{\RK}{R(K)}

\newcommand{\hi}{{i}}
\newcommand{\hj}{{j}}
\newcommand{\hk}{{k}}
\newcommand{\hl}{{l}}

\newcommand{\Tr}{\mathrm{Tr }}
\newcommand{\Tmult}{T}

\begin{document}
%%%%%%%%%%%%%%%%
%%%%%%%%%%%%%%%%

\begin{titlepage}
\begin{flushright}
September, 2017 \\
\end{flushright}
\vspace{5mm}

\begin{center}
{\Large \bf Non-conformal supercurrents in six dimensions
}
\\ 
\end{center}

\begin{center}

{\bf
Sergei M. Kuzenko${}^{a}$, Joseph Novak${}^{b}$ and Stefan Theisen${}^{b}$
} \\
\vspace{5mm}

\footnotesize{
${}^{a}${\it School of Physics M013, The University of Western Australia\\
35 Stirling Highway, Crawley W.A. 6009, Australia}}  
~\\
\vspace{2mm}
\footnotesize{
${}^{b}${\it Max-Planck-Institut f\"ur Gravitationsphysik, Albert-Einstein-Institut\\
Am M\"uhlenberg 1, D-14476 Golm, Germany}
}
~\\
\vspace{2mm}

\texttt{sergei.kuzenko@uwa.edu.au, joseph.novak@aei.mpg.de,  stefan.theisen@aei.mpg.de}\\
\vspace{2mm}

\end{center}

\begin{abstract}

Non-conformal supercurrents in six dimensions are described, which contain the 
trace of the energy-momentum tensor and the gamma-trace of the supersymmetry current amongst 
their component fields. Within the superconformal approach to $\cN = (1, 0)$ supergravity, 
we present various distinct non-conformal supercurrents, one of which is associated with an 
$\cO(2)$ (or linear) multiplet 
compensator, while another with a tensor multiplet compensator. We also 
derive an infinite class of non-conformal supercurrents involving $\cO(n)$ multiplets 
with $n > 2$. As an illustrative example we construct the relaxed hypermultiplet in 
supergravity. Finally, we put forward a non-conformal 
supercurrent in the $\cN = (2, 0)$ supersymmetric case.

\end{abstract}

\vfill

\vfill
\end{titlepage}

\newpage
\renewcommand{\thefootnote}{\arabic{footnote}}
\setcounter{footnote}{0}

\tableofcontents

\allowdisplaybreaks

%%%%%%%%%%%%%%%%%%%%%%%%%%%%%%%%%%%%%%%%%%%%%%%%%%%%%%
%%%%%%%%%%%%%%%%%%%%%%%%%%%%%%%%%%%%%%%%%%%%%%%%%%%%%%

\section{Introduction}

In supersymmetric field theory, the energy-momentum tensor belongs to
a supermultiplet, called the supercurrent \cite{FZ}. 
 In the case of superconformal theories, 
the supercurrent multiplet is irreducible. It contains the energy-momentum tensor $T_{mn}$, the spinor $Q$-supersymmetry current $S_m$ and the $R$-symmetry current $j_m$, 
in conjunction with some additional components that are required 
in order to have an equal number of bosonic and fermionic 
components.\footnote{Those conserved currents, which correspond to the other continuous transformations in the superconformal group, are constructed from the conformal supercurrent and  conformal Killing supervector fields \cite{KKT}.}

For theories without superconformal symmetry, the supercurrent multiplet is
reducible, as a rule.\footnote{There exist counter-examples in five and six dimenisons \cite{HL,HST83}.} 
The point is that the energy-momentum tensor is no longer traceless, 
and its trace $T^m{}_m$ may belong to a smaller supermultiplet embedded in
the non-conformal supercurrent. 
This trace supermultiplet 
also contains the $\g$-trace of the $Q$-supersymmetry current, $\g^m S_m$, 
as well as the divergence of the $R$-symmetry current, $\pa^m j_m$, 
if the $R$-symmetry current is not conserved.

An example worth recalling is the supercurrent multiplet  \cite{FZ} corresponding to 
$\cN=1$ Poincar\'e supersymmetry in four dimensions (4D). 
The conformal supercurrent is described by a real  axial vector superfield, 
$J_{m} = \bar J_m$,  
constrained by 
\bea
\bar D^\ad J_{\a\ad} =0~.
\eea
The non-conformal supercurrent proposed by Ferrara and Zumino \cite{FZ} is 
\bea
\bar D^\ad J_{\a\ad} = D_\a X~, \qquad \bar D_\ad X =0~.
\label{1.2}
\eea
Here $X$ is the trace supermultiplet.\footnote{Since 
$D^2 X - \bar D^2 \bar X = -2\ri \pa_{\a\ad} J^{\ad \a}$, 
the chiral scalar $X$ in \eqref{1.2} is an example of the 
so-called three-form multiplet \cite{Gates} (see \cite{GGRS} for a review). For instance, in quantum supersymmetric  Yang-Mills theories it holds that 
$\langle X \rangle = \k \,\tr (W^\a W_\a)$, 
where $k$ is a real parameter, and $W_\a$ the chiral field strength of the Yang-Mills supermultiplet. }

Unlike the conformal supercurrent, its non-conformal counterpart is not unique. 
The reason for this is that there may exist several inequivalent trace supermultiplets
supported by different supersymmetric field theories \cite{CPS}. For instance, 
another example of 4D $\cN=1$ non-conformal supercurrents 
is \cite{CPS,new,GGS}  
\bea
\bar D^\ad J_{\a\ad} = \c_\a ~, \qquad \bar D_\bd \c_\a =0~, 
\qquad 
D^\a \c_\a = \bar D_\ad \bar \c^\ad~.
\label{1.3}
\eea
Here the trace supermultiplet $\c_\a$ is a vector multiplet.

Similar to the energy-momentum tensor, which is the source of gravity,
the supercurrent is the source of supergravity \cite{OS,FZ2,Siegel77}.
For a given Poincar\'e supergravity theory, there often exist several off-shell 
formulations leading to the same dynamical system on shell.  However, different off-shell formulations for supergravity lead to different non-conformal supercurrents. In the case of 4D $\cN=1$ supergravity, 
for instance, the supercurrent multiplet \eqref{1.2} is associated with 
the old minimal formulation \cite{old}, while the conservation equation
\eqref{1.3} corresponds to the new minimal formulation \cite{new}.

The connection between the non-conformal supercurrents and different off-shell formulations 
for supergravity becomes more apparent in the Weyl-invariant (or conformal) approach to supergravity. 
Before discussing the supersymmetric case, 
it is instructive to recall 
the Weyl-invariant formulation for gravity.
Consider a system of matter fields $\vf^i$ coupled to the gravitational field.
In the approach of  \cite{Deser,Zumino,Dirac}, 
the gravitational field is described by the 
metric $g_{mn}$ and the conformal compensator $\f$, 
the latter being a nowhere vanishing scalar field.\footnote{As in     \cite{Deser,Zumino,Dirac}, our discussion here is restricted to the 4D case, but generalisation to higher dimensions is obvious.} 
The action must be invariant under general coordinate and Weyl transformations, 
\bea
\d g_{mn}=\nabla_m \l_n + \nabla_n \l_m - 2 \s g_{mn}~, \qquad 
\d \f =\l^m \nabla_m \f +  \s \f~,
\label{1.1}
\eea
augmented by certain transformations of the matter fields. Consider the matter action
\bea 
S_{\rm M} = \int \rd^{4}x \, \sqrt{-g} \, \cL( \vf^i ; g_{mn}, \f)  \ .
\eea
If the metric and the compensator acquire arbitrary infinitesimal 
displacements, $g_{mn} \to g_{mn} +\d g_{mn}$ and $\f \to \f +\d \f$, 
the action  varies as 
\bea
\d S_{\rm M} = \int \rd^{4}x \, \sqrt{-g} \, \Big\{ \hf T^{mn}\d g_{mn} + T \d \ln \f \Big\}~,
\eea
where $T^{mn}$ denotes the energy-momentum tensor of the system. 
If the matter fields are chosen to obey their equations of motion, 
$\d S_{\rm M} / \d \vf^i =0$, the conditions of invariance of $S_{\rm M}$ under the local transformations \eqref{1.1} are
\begin{subequations}
\bea
\nabla_n T^{mn} &=& T \nabla^m \ln \f~, \label{1.4a.}\\
g_{mn} T^{mn} &=&T~.
\eea
\end{subequations}
The Weyl invariance may be used to impose a condition $\f ={\rm const}$, 
in which case eq. \eqref{1.4a.} turns into
\bea
\nabla_n T^{mn} =0~,
\eea
which is the standard conservation equation.

In analogy with the Weyl-invariant formulation for gravity \cite{Deser,Zumino,Dirac}, Poincar\'e or anti-de Sitter supergravity theories 
may be formulated as conformal supergravity coupled to a compensating supermultiplet \cite{Siegel77_2,KakuT}. Unlike gravity, however, 
supergravity generally allows for several 
choices of conformal compensator that differ in their auxiliary fields. 
It turns out that 
different conformal compensators lead to different off-shell supergravity theories and, as a consequence, to different supercurrent multiplets. For instance, the conservation equation \eqref{1.2} of the old 
minimal formulation of supergravity corresponds to a compensating chiral scalar multiplet, while 
the conservation equation \eqref{1.3} of the new minimal formulation of supergravity corresponds 
to a compensating tensor multiplet.

For 6D $\cN=(1,0)$ supersymmetry, the conformal supercurrent 
was described more than thirty years ago \cite{HST83}. However,
to the best of our knowledge,  
no classification of non-conformal supercurrents has been given. 
The only known non-conformal $\cN =(1,0) $ supercurrent was proposed 
by Manvelyan and R\"uhl  \cite{MR03}. 
It proves to be a 6D analogue of the 4D $\cN=2$ 
 non-conformal supercurrent introduced by Stelle \cite{Stelle}. 
 The latter obeys the conservation equation
 \bea
 {\bar D}^{ij} J =  \frac{1}{5} {\bar D}_{kl} {{\mathbb L}}^{klij} ~, 
 \qquad \bar D^{ij} := \bar D_{\ad}^i \bar D^{\ad j}~,
\label{1.4}
\eea
where $J= \bar J$ denotes the $\cN=2$ supercurrent \cite{Sohnius,HST}.
The trace supermultiplet $ {{\mathbb L}}^{ijkl} ={\mathbb L}^{(ijkl)} $ is real, 
$\overline{{\mathbb L}^{ijkl}} = {\mathbb L}_{ijkl}$, 
and is subject to the analyticity constraints defining an $\cO(4)$ multiplet,
\bea
D^{(i}_\a  {{\mathbb L}}^{jklm)} = {\bar D}^{(i}_\ad  {{\mathbb L}}^{jklm)}=0~.
\eea   
It was shown in \cite{KT,BK10} that the conservation equation \eqref{1.4} 
naturally occurs in theories which couple to the $\o$-hypermultiplet 
compensator \cite{GIKOS,GIOS87} within the harmonic superspace 
approach to 4D $\cN=2$ supergravity  (see \cite{GIOS} for a review). 

The purpose of this paper is twofold: (i) to derive various consistent 6D  
non-conformal supercurrents; and (ii) to lift them to an arbitrary curved conformal 
supergravity background with a conformal compensator. As a consequence, all 
non-conformal supercurrents may be classified by the choice of compensating conformal 
supermultiplet.

This paper is organised as follows. In section \ref{(1,0)AnomSupercurrent} we 
present an infinite family of 6D $\cN = (1,0)$ non-conformal supercurrents 
involving $\cO(n)$ multiplets for $n > 1$ and we 
illustrate a couple of them by analysing the equations of motion for certain models. 
Section \ref{dilWeylmult} is devoted to the special case of using an $\cN = (1,0)$ tensor multiplet 
as a compensator. We put forward a non-conformal supercurrent for the $\cN = (2, 0)$ 
case in section \ref{(2,0)supercurrent}. Finally, we discuss our results in section \ref{discussion}.

We have included a number of technical appendices. Throughout this paper 
we will make use of the superspace formulation of conformal supergravity known 
as 6D $\cN = (1,0)$ conformal superspace \cite{BKNT}. Therefore, we provide the 
salient details of conformal superspace in appendix \ref{geometry}. Appendix \ref{PrepotO(2)}
is devoted to the prepotential description of the $\cO(2)$ (or linear) multiplet. In 
appendix \ref{YMmultiplet}, we summarise the description of the Yang-Mills multiplet 
in conformal superspace. Finally, we give a superform description of the $\cN = (2, 0)$ 
tensor multiplet and its deformation in appendix \ref{c4Form}.

%%%%%%%%%%%%%%%%%%%%%%%%%%%%%%%%%%%%%%%%%%%%%%%%%%%%%%
%%%%%%%%%%%%%%%%%%%%%%%%%%%%%%%%%%%%%%%%%%%%%%%%%%%%%%

\section{Non-conformal $\cN = (1, 0)$ supercurrents} 
\label{(1,0)AnomSupercurrent}

In 6D  $\cN = (1, 0)$ supergravity, the conformal supercurrent $J$
is a primary superfield of dimension $+4$,
\be \mathbb D J = 4 J \ , \quad S^\a_i J = 0 \ ,
\ee
which satisfies the conservation equation \cite{BKNT} 
\be \nabla^{\a ijk} J = 0 \ , \quad \nabla^{\a ijk} := \frac{1}{3!} \eps^{\a\b\g\d} \nabla_\b^{(i} \nabla_\g^j \nabla_\d^{k)} \ . \label{ConformConservEqn}
\ee
In the flat superspace limit, this equation reduces to the one originally given in \cite{HST}. 
 
In the presence of a conformal compensator the conservation equation 
\eqref{ConformConservEqn}
is deformed to
\be \nabla^{\a ijk} J = A^{\a ijk} \ , \label{consEqAnomalous}
\ee
where $A^{\a ijk}$ is a primary superfield of dimension $\frac{11}{2}$. Using 
the identity
\be \nabla_\a^{(i} \nabla^{\b jkl)} = \frac{1}{4} \d_\a^\b \nabla_\g^{(i} \nabla^{\g jkl)} \ ,
\ee
it can be checked that $A^{\a ijk}$ must 
satisfy the following integrability condition:
\be \nabla_\a^{(i} A^{\b jkl)} = \frac{1}{4} \d^\b_\a \nabla_\g^{(i} A^{\g jkl)} \ . \label{anomalyConsistEq}
\ee

In order to guarantee the existence of a {\it conserved} 
supersymmetry current and energy-momentum tensor, 
the integrability condition \eqref{anomalyConsistEq} has to be accompanied
with some additional requirements on the structure of $A^{\a ijk}$.
To understand this in more detail, it is worth analysing the 
deformed conservation equation \eqref{consEqAnomalous} 
in Minkowski superspace and uncover the corresponding component structure. 
In what follows, we will refer to the superfield $A^{\a ijk}$ as the {\it trace superfield} 
since in general it gives a trace contribution to the energy-momentum tensor, 
while $J$ only contains a symmetric traceless contribution.

%%%%%%%%%%%%%%%%%%%%%%%%%%%%%%%%%%%%%%%%%%%%%%%%%%%%%%

\subsection{Non-conformal supercurrents in Minkowski superspace} \label{MinkAnalysSection}

In this subsection we will make use of the spinor derivatives
for 6D $\cN=(1,0)$ Minkowski superspace, $D_\a^i$, which satisfy the 
anti-commutation relation
\be \{ D_\a^i , D_\b^j \} = - 2 \ri \eps^{ij} \partial_{\a\b}
\ee
and commute with partial vector derivatives, $[\partial_a , D_\a^i] = 0$.

We now analyse the component structure of the superfields $J$ and 
$A^{\a ijk}$ subject to the general constraints \eqref{consEqAnomalous} and 
\eqref{anomalyConsistEq} in the flat-superspace case.\footnote{In the anomaly-free
case, $A^{\a ijk}=0$, the component analysis was carried out in \cite{HST83}.
More recently, it was generalised \cite{MR03} to the case of a special trace 
supermultiplet $A^{\a ijk}$ given by \eqref{MRconstruction}.}
Taking successive spinor derivatives of the trace superfield $A^{\a ijk}$, 
one finds\footnote{The $\rm SU(2)$ indices 
on any field are always assumed to be symmetrized.}
\bsubeq
\bea 
D_\a^i A^{\b jkl} &=& \d_\a^\b A^{ijkl} + \eps^{i(j} A_\a{}^\b{}^{kl)} + \d_\a^\b \eps^{i(j} A^{kl)} \ , \quad 
A_\a{}^\a{}^{ij} = 0 \ , \label{Descendents2.6a} \\
D_\a^p A^{ijkl} &=& \eps^{p(i} \L_\a{}^{jkl)} \ , \\
D_\a^{k} A^{ij} &=& \frac{\ri}{2} \partial_{\a\b} A^{\b ijk} + \eps^{k(i} \L_\a^{j)} \ , \\ 
D_\a^{k} A_\b{}^{\g ij} &=& - \d_\a^\g \L_\b{}^{ijk} + \frac{1}{4} \d_\b^\g \L_\a{}^{ijk}
- \frac{4}{3} \eps^{k(i} \d^\g_{\a} \L_{\b}^{j)}
+ \frac{1}{3} \eps^{k(i} \d^\g_{\b} \L_{\a}^{j)}
+ \eps^{k(i} \L_{\a\b}{}^{\g j)} \non\\
&&+ 2 \ri \partial_{\a\b} A^{\g ijk} - \frac{\ri}{2} \d^\g_\b \partial_{\a\d} A^{\d ijk} 
\ , \quad \L_{\a\b}{}^{\b ij} = 0 \ , \quad \L_{\a\b}{}^{\g ij} = \L_{[\a\b]}{}^{\g ij} \ , \\
D_\a^{l} \L_\b{}^{ijk} &=& 
2 \ri \partial_{\a\b} A^{ijkl}
+ \eps^{l(i} \cA_{\a\b}{}^{jk)}
\ , \quad \cA_{\a\b}{}^{ij} = \cA_{[\a\b]}{}^{ij} \ , \\
D_\a^{i} \L_\b^j &=& 
\eps^{ij} \cA_{\a\b} 
+ \frac{\ri}{2} \partial_{\g\a} A_\b{}^{\g}{}^{ij} 
- \frac{\ri}{6} \partial_{\g\b} A_\a{}^\g{}^{ij}
+ \frac{4 \ri}{3} \partial_{\a\b} A^{ij}
\ , \quad \cA_{\a\b} = \cA_{[\a\b]} 
\ , \\
D_\a^{i} \L_{\b\g}{}^{\d j} &=& 
\frac{2}{3} \d^\d_{[\b} \cA_{\g]\a}{}^{ij}
+ \d^\d_\a \cA_{\b\g}{}^{ij}
+ 4 \ri \partial_{\a [\b} A_{\g]}{}^{\d}{}^{ij}
+ \frac{2 \ri}{3} \partial_{\b\g} A_\a{}^\d{}^{ij} \non\\
&&+ \frac{4 \ri}{9} \d^\d_{[\b} \partial_{\g] \r} A_\a{}^\r{}^{ij}
- \frac{4 \ri}{3} \d_{[\b}^\d \partial_{|\a\r|} A_{\g]}{}^{\r ij} 
\non\\
&&+ \frac{4 \ri}{9} \d^\d_{[\b} \partial_{\g]\a} A^{ij}
+ \frac{2 \ri}{3} \d^\d_\a \partial_{\b\g} A^{ij}
\non\\
&&
+ \eps^{ij} \eps_{\a\b\g\r} \cS^{\r \d} + \frac{4}{3}\eps^{ij} \d^\d_{[\b} \cA_{\g]\a} + 2 \eps^{ij} \d^\d_\a \cA_{\b\g}
\ , \quad \cS^{\a\b} = \cS^{(\a\b)}
\ .
\eea
\esubeq
Taking 
successive spinor covariant derivatives of the superfield $J$ satisfying the equation 
\eqref{consEqAnomalous}, one finds
the following relations
{\allowdisplaybreaks
\bsubeq
\bea 
D_\a^i J &=& \J_\a^i \ , \\
D_\a^i \J_\b^j &=& V_{\a\b}{}^{ij} + \eps^{ij} C_{\a\b} - \ri \eps^{ij} \partial_{\a\b} J \ , \quad C_{\a\b} = C_{(\a\b)} \ , \quad V_{\a\b}{}^{ij} = V_{[\a\b]}{}^{ij} \ , 
\\
D_\a^i C_{\b\g} &=& 
\S^i_{(\b}{}_{, \g) \a} 
+ \frac{8 \ri}{5} \partial_{\a(\b} \J_{\g)}^{i}
\ , \quad \S_{[\a}^i{}_{, \g\d]} = 0 \ , \\
D_\a^i V_{\b\g}{}^{jk}
&=& \eps^{i (j} \S_\a^{k)}{}_{, \b\g}
- \frac{8 \ri}{5} \eps^{i(j} \partial_{\a[\b} \Psi_{\g]}^{k)}
- \frac{2 \ri}{5} \eps^{i(j} \partial_{\b\g} \Psi_{\a}^{k)}
- \eps_{\a\b\g\d} A^{\d ijk}
\ , \\
D_\a^i \S_\b^j{}_{, \g\d} &=& 
\eps^{ij} \cT_{\a\b , \g\d}
+ \frac{4 \ri}{3} \partial_{\a\b} V_{\g\d}{}^{ij} - \frac{4 \ri}{15} \partial_{\g\d} V_{\a\b}{}^{ij}
- \frac{4 \ri}{15} \partial_{\b[\g} V_{\d]\a}{}^{ij}
- \frac{4 \ri}{3} \partial_{\a[\g} V_{\d]\b}{}^{ij}
\non\\
&&
+ 2 \ri \eps^{ij} \partial_{\a[\g} C_{\d]\b}
- \frac{2 \ri}{5} \eps^{ij} \partial_{\b [\g} C_{\d]\a}
+ \frac{2 \ri}{5} \eps^{ij} \partial_{\g\d} C_{\a\b} 
\non\\
&&
- \eps_{\g\d\e\a} A_\b{}^\e{}^{ij}
+ \frac{1}{3} \eps_{\g\d\e\b} A_\a{}^\e{}^{ij}
\ , \quad \cT_{\a[\b , \g\d]} = 0 \ , \quad \cT_{\a\b , \g\d} = \cT_{[\a\b] , [\g\d]} \ , ~~~~\label{TindetitiesDer} \\
D_\e^i \cT_{\a\b , \g\d} &=&
\frac{2 \ri}{3} \partial_{[\a[\g} \S_{|\e|}^i{}_{, \d]\b]}
+ 2 \ri \partial_{\e[\a} \S_{\b]}^i{}_{, \g\d}
+ 2 \ri \partial_{\e[\g} \S_{\d]}^i{}_{, \a\b} \non\\
&&+ \frac{\ri}{3} \partial_{\a\b} \S_\e^i{}_{ , \g\d}
+ \frac{\ri}{3} \partial_{\g\d} \S_\e^i{}_{, \a\b}
+ \hf \eps_{\a\b\r\e} \L_{\g\d}{}^{\r i}
+ \hf \eps_{\g\d\r\e} \L_{\a\b}{}^{\r i}
\ ,
\eea
\esubeq
}
as well as the conditions
\bea
\partial^{\a\b} V_{\a\b}{}^{ij} &=& 4 \ri A^{ij} \ , \quad \partial^{\a\b} \S_\g^i{}_{, \a\b} = 4 \ri \L_\g^i \ , \quad
\partial^{\a\b} \cT_{\a\b , \g\d} = 4 \ri \cA_{\g\d} \ . \label{VSTconserve}
\eea
Note that 
the algebraic properties of the tensor  $\cT_{\a\b , \g\d}$, which are 
given in eq. \eqref{TindetitiesDer}, imply the identity
\bea
\cT_{\a\b , \g\d} = \cT_{\g\d , \a\b}~.
\eea
As a result, if we convert each of the two pairs of spinor indices of $\cT_{\a\b , \g\d} $ 
into vector ones by the standard rule $V_{\a\b} =-V_{\b\a}\to V_a = \frac{1}{4} (\tilde{\g}_a)^{\a\b} V_{\a\b} $, 
we end up with a second-rank tensor
$\cT_{a b}$, which is symmetric and traceless, $\cT_{a b}=\cT_{ba }$
and $ \cT^a{}_a=0$.

The equations \eqref{VSTconserve} tell us that if $A^{\a ijk} = 0$ the 
component projections of $V_{\a\b}{}^{ij}$, 
$\S_\g^i{}_{, \a\b}$ and $\cT_{\a\b , \g\d}$ are proportional to the conserved SU(2) current, 
supersymmetry current and energy-momentum 
tensor, respectively.\footnote{One can also verify that the supercurrent has 
$40+40$ component degrees of freedom.} 
If an arbitrary trace superfied $A^{\a ijk}$ is switched on, 
they are no longer conserved. It follows from \eqref{VSTconserve} 
that in order to be able to specify a conserved supersymmetry current 
it is necessary to require that $\L_\a^i$ is a vector divergence, $\L_\a^i = \partial^b \tilde{\S}_\a^i{}_{,b}$.
It is now important to note  that if $\L_\a^i$ is a divergence then so is $\cA_{\a\b}$, i.e. 
a conserved supersymmetry current automatically implies a conserved energy-momentum tensor.  
Similarly, a conserved $\rm SU(2)$ current implies both a conserved supercurrent and a 
conserved energy-momentum 
tensor. 
One should, however, keep in mind that 
the \emph{conserved} supersymmetry current and energy-momentum 
tensor need no longer be $\g$-traceless and 
traceless, respectively. 

Let us see how this works for the non-conformal supercurrent involving 
an $\cO(4)$ multiplet \cite{MR03}. 
The trace superfield $A^{\a ijk}$ is chosen to be proportional to
\be A^{\a ijk} = \ri \partial^{\a\b} D_{\b l} {\mathbb L}^{ijkl} \ , \qquad
\label{MRconstruction}
\ee
where ${\mathbb L}^{ijkl} = {\mathbb L}^{(ijkl)}$ satisfies the reality condition 
$\overline{{\mathbb L}^{ijkl} }={\mathbb L}_{ijkl} $ and the defining constraint 
for an $\cO(4)$ multiplet 
\be D_\a^{(i} {\mathbb L}^{jklp)} = 0 \ .
 \ee
It is simple to check that this superfield satisfies the integrability condition \eqref{anomalyConsistEq} in 
the flat case. Furthermore, since the trace multiplet \eqref{MRconstruction} 
is a divergence, its descendent
$A^{ij}$ is a divergence and a conserved SU(2) current can be introduced. As remarked above, 
it then follows that a conserved $Q$-supersymmetry current and 
energy-momentum tensor exist as well. These currents may be defined as follows:
\bsubeq \label{O(4)constructionCurrents}
 \begin{align}
 j_{\a\b}{}^{ij} &= V_{\a\b}{}^{ij} + \frac{3}{4} D_{\a k} D_{\b l} {\mathbb L}^{ijkl} \ , \quad \partial^{\a\b}  j_{\a\b}{}^{ij} = 0 \ , \\
 S_{\a\b ,}{}_\g^i &= \S_\g^i{}_{,\a\b} + \hf D_{\g j} D_{[\a k} D_{\b] l} {\mathbb L}^{ijkl} \ , \quad \partial^{\a\b} S_{\a\b ,}{}_\g^i = 0 \ , \\
 T_{\a\b , \g\d} &= \cT_{\a\b , \g\d} + \frac{1}{4} D_{[\a i} D_{\b j} D_{\g k} D_{\d] l} {\mathbb L}^{ijkl} \ , \quad \partial^{\a\b}  T_{\a\b , \g\d} = 0 \ .
 \end{align}
 \esubeq
Note that neither is the $Q$-supersymmetry current $\g$-traceless nor is  the energy-momentum tensor traceless \cite{MR03}.

Within the conformal approach, the form of the trace superfield $A_\a{}^{ijk}$ should depend on conformal compensators. 
Therefore, a  natural question one can ask is: what compensator(s) should one associate with 
the construction \eqref{MRconstruction}? Furthermore, how do we lift the construction \eqref{MRconstruction} to a primary 
superfield with the use of a compensator in conformal supergravity? One can show that if we assume that the compensator is 
a tensor multiplet and we try to lift the construction \eqref{MRconstruction} to conformal superspace (see appendix \ref{geometry}), 
it is not possible 
to add compensator dependant terms such that the condition \eqref{anomalyConsistEq} is satisfied. On the other hand, if the 
compensator was an $\cO(2)$ (or linear) multiplet one would expect a symmetric SU(2) tensor to appear in 
the construction \eqref{MRconstruction}, which is not the case.\footnote{The SU(2) tensor corresponding to the 
superfield describing the $\cO(2)$ multiplet can be set to a constant using super-Weyl transformations.} 
For this reason, it is necessary to use a different scalar compensating 
superfield instead that of the tensor multiplet. We 
will present the appropriate compensator and show how to generalise the construction in \cite{MR03} to supergravity in section \ref{anomSuperfieldO(4)2.3}.

It is elucidating to ask what can be learned by allowing the $\cO(4)$ 
multiplet ${\mathbb L}^{ijkl}$ to be composite. 
For instance, suppose we have two $\cO(2)$ multiplets 
described by the superfields $G^{ij} = G^{(ij)}$ and $H^{ij} = H^{(ij)}$, which satisfy the differential constraints
\be D_\a^{(i} G^{jk)} = D_\a^{(i} H^{jk)} = 0 \ .
\ee
We can then construct 
\be {\mathbb L}^{ijkl} = G^{(ij} H^{kl)} \ .
\ee
We will further assume $G^{ij}$ has a nowhere vanishing magnitude $G \neq 0$, which is defined by $G^2 := \hf G^{ij} G_{ij}$. 
If we freeze $G$ to a constant we find
\be 0 = D_\a^i G^2 = \frac{2}{3} G^{ij} D_\a^k G_{jk} \ \implies \ D_\a^i G^{jk} = \frac{2}{3} \eps^{i(j} D_{\a k} G^{j)k} = 0 \label{(2.12)}
\ ,
\ee
where we used $G^{ij} G_{jk} = \d^i_k G^2$.
Using \eqref{(2.12)}, the superfield $A^{\a ijk}$ can be seen to take the form
\be A^\a{}^{ijk} = \ri G^{(ij} {\mathbb W}^{\a k)} \ , \label{flatcasePostAnomaly}
\ee
where we have defined
\be {\mathbb W}^{\a i} := \frac{5}{6} \partial^{\a\b} D_{\b j} H^{ij} \ . \label{flatCompositeVMO(2)}
\ee
One can verify that ${\mathbb W}^{\a i}$ satisfies the following
differential constraints:
\bsubeq
\bea D_\a^{(i} {\mathbb W}^{\b j)} &=& \frac{1}{4} \d_\a^\b D_\g^{(i} {\mathbb W}^{\g j)} \ , \label{VMbbWconsta} \\
D_{\a i} {\mathbb W}^{\a i} &=& 0 \label{VMbbWconstb} \ ,
\eea
\esubeq
which correspond to those of a vector multiplet, see e.g. appendix \ref{YMmultiplet}.

It is important to point out that the representation \eqref{flatcasePostAnomaly} actually implies 
the existence of a conserved supersymmetry current and energy-momentum tensor due to the 
constraints \eqref{VMbbWconsta} and \eqref{VMbbWconstb}, irrespective of the 
form \eqref{flatCompositeVMO(2)}. In other words, we only need to require ${\mathbb W}^{\a i}$ to 
be an off-shell vector multiplet for these currents to exist. In particular, the constraint \eqref{VMbbWconsta} 
implies the condition \eqref{anomalyConsistEq}, while the constraint \eqref{VMbbWconstb} is 
required to show\footnote{Keep in mind that $G^{ij}$ is constant.}
\bea
\L_\a^i =  \frac{1}{8} D_{\a j} D_{\b k} A^{\b ijk}
= \frac{2}{3} \partial_{\a\b} ( G^{ij} {\mathbb W}^{\b}_j ) \ ,
\eea
which ensures the existence of a conserved supersymmetry current that is, however, no longer $\gamma$-traceless, as well 
as a conserved energy-momentum tensor that is no longer traceless. These may be defined as
\bsubeq
\begin{align}
S_{\a\b ,}{}_\g^i &:= \S_\g^i{}_{, \a\b} - \frac{4 \ri}{3} \eps_{\a\b\g\d} G^{ij} {\mathbb W}^\d_j \ , \quad \partial^{\a\b} S_{\a\b ,}{}_\g^i = 0 \ , \\
T_{\a\b , \g\d} &:= \cT_{\a\b , \g\d} + \frac{\ri}{6} \eps_{\a\b\g\d} G^{ij} D_{\e i} {\mathbb W}^{\e}_j \ , \quad \partial^{\a\b}  T_{\a\b , \g\d} = 0 \ .
\end{align}
\esubeq
The off-shell conditions \eqref{VMbbWconsta} and \eqref{VMbbWconstb} do not lead 
to an $\rm SU(2)$ current, since one finds
\be A^{ij} = \frac{\ri}{8} G^{k(i} D_\a^{j)} {\mathbb W}^{\a}_k \ .
\ee
However, as $G^{ij} A_{ij} = 0$ we can instead introduce a conserved U(1) current $j_{\a\b}$ defined by
\be j_{\a\b} = V_{\a\b}{}^{ij} G_{ij} \ , \quad \partial^{\a\b} j_{\a\b} = 0 \ ,
\ee
where the U(1) subgroup is the stability group of $G^{ij}$. 
Only when the equations of motion require $D_\a^{(i} {\mathbb W}^{\a j)}$ 
to be a vector divergence up to terms proportional to $G^{ij}$ can a conserved 
SU(2) current be introduced.

We conclude this section by emphasising once more that eq. \eqref{flatcasePostAnomaly} 
leads to a non-conformal supercurrent for any vector multiplet ${\mathbb W}^{\a i}$. 
The appearance of the 
constant SU(2) tensor $G^{ij}$ is to be interpreted as a compensator that has been frozen. 
It is precisely the form \eqref{flatcasePostAnomaly} that we will generalise to curved superspace 
in the next subsection and it will be verified by further analysis and a worked example.

%%%%%%%%%%%%%%%%%%%%%%%%%%%%%%%%%%%%%%%%%%%%%%%%%%%%%%

\subsection{The non-conformal supercurrent based on a compensating $\cO(2)$ multiplet} \label{anomSuperfieldO(2)}

Let us first describe how the conservation condition 
on the supercurrent originates from superspace. We will consider a supergravity action 
with a compensating $\cO(2)$ multiplet $G^{ij}$. The supergravity equations 
of motion 
in superspace can be easily obtained if one knows the dependence of 
the supergravity action on the unconstrained superfield prepotential for 
$\cN = (1, 0)$ conformal supergravity. It is a real primary scalar $H$ \cite{HST83} of dimension $-2$ 
with supergravity gauge transformation\footnote{The 
gauge transformation presented here is the unique extension of the linearised transformation in \cite{HST83} 
to curved superspace, assuming the gauge parameter $\L_{\a ijk}$ is primary.}
\be \d H = \nabla^{\a ijk} \L_{\a ijk} \ , \quad \L_{\a ijk} = \L_{\a (ijk)} \ . \label{SUGRAtrans}
\ee
In general, the constrained superfields must also transform under such a gauge transformation since their
constraints must be preserved under shifts in the supergravity prepotential. This tells us 
that under the gauge transformation \eqref{SUGRAtrans} the prepotential for the $\cO(2)$ 
multiplet $\r_{\a i}$ should transform. For a description of the $\cO(2)$ multiplet in terms of 
its prepotential in supergravity see appendix \ref{PrepotO(2)}.

If any additional matter fields other than the compensator are chosen to obey their equations of motion, a 
general variation of the action with respect to the supergravity and compensator prepotentials becomes
\be \d S = \int \rd^{6|8}z \, E \, \Big( \d H \, J +  \d \r_{\a }^i \, {\mathbb W}^{\a }_i  \Big) \ .
\label{deltaS}
\ee
The  prepotential $\r_\a^i$ is defined modulo gauge transformations
\be \r_\a^i \rightarrow \r'{}_\a^i = \r_\a^i + \nabla_\a^i \t + \nabla_{\b j} \t_\a{}^\b{}^{ij} \ , \quad \t_\a{}^\a{}^{ij} = 0 \ , \quad 
\t_\a{}^{\b}{}^{ij} = \t_\a{}^{\b}{}^{(ij)} \ , 
\ee
where $\t$ and $\t_\a{}^\b{}^{ij}$ are dimensionless primary superfields. 
In order for the action to be invariant under 
these gauge transformations, 
the superfield ${\mathbb W}^{\a }_i$ 
 must obey the constraints
 \bea 
 \nabla_\a^{(i} {\mathbb W}^{\b j)} &=& \frac{1}{4} \d_\a^\b \nabla_\g^{(i} {\mathbb W}^{\g j)} \ , \qquad
\nabla_{\a i} {\mathbb W}^{\a i} = 0 \ ,
\eea
which are characteristic of the field strength of  a vector multiplet.  
Now we wish to specialise to the supergravity gauge transformations where $\d S = 0$ but we 
need to know the transformation of the prepotential $\r_{\a i}$. Its transformation should involve the 
supergravity gauge parameter $\Lambda_{\alpha ijk}$ and covariant fields of the compensating $\cO(2)$ multiplet
since we should obtain a covariant conservation equation.\footnote{The transformation must also 
be linear in the fields of the $\cO(2)$ multiplet 
since its prepotential description must remain unchanged. We will also verify 
this by an explicit example.} 
On dimensional grounds we must have\footnote{We can always rescale $G^{ij}$ by fixing its relative normalisation to its prepotential.}
\be \d \r_{\a i} = \ri \L_{\a ijk} G^{jk} \ .
\ee
Requiring $\d S = 0$ under the supergravity gauge transformations leads 
to the non-conformal conservation equation
\be \nabla^{\a ijk} J = \ri G^{(ij} {\mathbb W}^{\a k)}  \ , 
\label{convEqO(2)AnomalyA}
\ee
where the $\cO(2)$ multiplet in conformal superspace is 
a primary superfield $G^{ij} = G^{(ij)}$ of dimension 4 satisfying the constraint
\be \nabla_\a^{(i} G^{jk)} = 0 \ .
\ee
Therefore, the general form of the trace superfield $A^{\a ijk}$ in the presence of a compensating 
nowhere vanishing $\cO(2)$ multiplet 
(with $G := \hf G^{ij} G_{ij} \neq 0$) in supergravity  is
\be A^{\a ijk} = \ri G^{(ij} {\mathbb W}^{\a k)} \ , \label{O(2)AnomalyA}
\ee
where ${\mathbb W}^{\a i}$ is the composite vector multiplet determined by the theory via
\eqref{deltaS}.

It is interesting to note that if we weaken the constraint defining the $\cO(2)$ multiplet 
to a deformed $\cO(2)$ multiplet \cite{KNS1, KNS2}
\be \nabla_\a^{(i} G^{jk)} = \ri \eps_{\a\b\g\d} {\mathbb W}^{\b (i} {\mathbb W}^{\g j} {\mathbb W}^{\d k)} \ ,
\ee
the postulated trace superfield \eqref{O(2)AnomalyA} still satisfies the consistency 
condition \eqref{anomalyConsistEq} but a conserved supersymmetry current is no longer guaranteed.

%%%%%%%%%%%%%%%%%%%%%%%%%%%%%%%%%%%%%%%%%%%%%%%%%%%%%%

\subsubsection{An example: Abelian gauge theory coupled to an $\cO(2)$ multiplet} \label{exampleO(2)}

We now provide an explicit example of a non-conformal supercurrent in curved superspace. Consider an 
Abelian gauge theory coupled to a linear multiplet. The action for the 
theory is built out of two supersymmetric invariants: (i) a higher-derivative Abelian vector multiplet action; and (ii)
a $BF$ action giving rise to the coupling of the vector multiplet to the linear (or $\cO(2)$) multiplet.

A supersymmetric $F \Box F$ action was described in \cite{ISZ05} in Minkowski superspace and in 
conformal supergravity in \cite{BKNT}. It is straightforward to construct its supercurrent (up to 
some normalisation). In the Abelian case, it is\footnote{We refer the reader to appendix \ref{YMmultiplet} 
for our notation and conventions regarding the vector multiplet.}
\be
J = \frac{3}{8} X^{ij} X_{ij} + \frac{\ri}{2} W^{\a i} \nabla_{\a\b} W^{\b}_i + \frac{1}{4} F_\a{}^\b F_\b{}^\a \ ,
\ee
while the equation of motion for the vector multiplet is
\be \cG^{ij} =0 \ , \qquad
\cG^{ij} := \Box X^{ij} - 2 Y_\a{}^\b{}^{ij} F_\b{}^\a 
+ \frac{5}{2} X^{\a (i} \overleftrightarrow{\nabla}_{\a\b} W^{\b j)} \ , \label{EoMFBoxF+BF}
\ee
where we have defined $\Box := \nabla^a \nabla_a$ and $S {\overleftrightarrow{\nabla}}_a U := S \nabla_a U - (\nabla_a S) U$ for arbitrary 
superfields $S$ and $U$. 
One can check that, upon using the equations of motion, the supercurrent is conserved
\be
\nabla^{\a ijk} J = \ri W^{\a (i} \cG^{jk)} = 0 \ .
\ee

The non-conformal conservation condition can be deduced 
if we consider the $F \Box F$ action coupled 
to a $BF$ action, where $B$ is the gauge four-form 
of a nowhere vanishing $\cO(2)$ multiplet. The $BF$ action is just the 
action formula for the product of a 
vector and an $\cO(2)$ multiplet \cite{GL85,BSVanP},
\be S_{BF} = \int \rd^{6|8}z \, E \, W^{\a i} \r_{\a i} \ .
\ee
We take the $\cO(2)$ multiplet 
as a conformal compensator. For the combined action 
incorporating the $F \Box F$ action and the $BF$ action, the equation of 
motion for the vector multiplet becomes
\be \cG^{ij} = \l G^{ij} \ ,
\label{eomG}
\ee
where $\l$ is some constant and $G^{ij}$ is the compensating $\cO(2)$ multiplet with prepotential $\rho_{\alpha i}$. 
It is straightforward to check that the non-conformal 
conservation equation \eqref{convEqO(2)AnomalyA} holds due to the equation of motion,
\be \nabla^{\a ijk} J = \ri G^{(ij} {\mathbb W}^{\a k)} \ , \quad {\mathbb W}^{\a i} = \l W^{\a i} \ ,
\ee
which verifies the supercurrent conservation equation for a compensating $\cO(2)$ multiplet. In the above computation it 
is important to note that the supercurrent for the combined action does not obtain a contribution from 
the $BF$ action since it does not depend on the supergravity prepotential as it admits a topological realisation.

Finally, it is worth mentioning that the equations of motion for this example lead to a 
conserved SU(2) current. Indeed, using  \eqref{eomG} and \eqref{EoMFBoxF+BF} we find, after 
reducing to flat superspace, 
\be A^{ij} = \frac{\l}{2} G^{k(i} X^{j)}{}_k = \frac{1}{2} (\Box X^{k(i}) X^{j)}{}_k = \frac{1}{2} \partial^a \Big( (\partial_a X^{k(i}) X^{j)}{}_k \Big) \ .
\ee
The conserved $\rm SU(2)$ current is therefore
\be j_{\a\b}{}^{ij} = V_{\a\b}{}^{ij} - \frac{\ri}{2} (\partial_{\a\b} X^{k(i}) X^{j)}{}_k \ , \quad \partial^{\a\b} j_{\a\b}{}^{ij} =0 \ .
\ee

%%%%%%%%%%%%%%%%%%%%%%%%%%%%%%%%%%%%%%%%%%%%%%%%%%%%%%

\subsection{The non-conformal supercurrent involving an $\cO(4)$ multiplet} \label{anomSuperfieldO(4)2.3}

The trace multiplet based on the $\cO(4)$ multiplet, 
eq. \eqref{MRconstruction}, must correspond to a conformal compensator 
described by a scalar superfield which cannot be the tensor multiplet as mentioned 
previously. 
It turns out that 
the right compensating multiplet is built from a primary 
dimension $-4$ scalar superfield $\Tmult$ subject to the 
constraint\footnote{Coupling this multiplet to 
conformal supergravity is equivalent to working in the SU(2) superspace formulation 
of \cite{LT-M12} and setting the torsion component $N_{\a\b} = 0$.}
\be \nabla^k_{(\a} \nabla_{\b) k} \Tmult = 0 \implies \nabla_{(\a}^i \nabla_{\b)}^j \Tmult = 0 \ .
\ee
It corresponds to the $\cO^{*}(4)$ multiplet described in \cite{KNT17} and 
the above constraint can be solved in terms of an unconstrained prepotential 
$U_{ijkl} = U_{(ijkl)}$ as\footnote{This was 
first worked out in Minkowski superspace in \cite{HST83}.}
\be \Tmult = \nabla^{ijkl} U_{ijkl} \ , \quad 
\nabla^{ijkl} := \frac{1}{4!} \eps^{\a\b\g\d} \nabla_\a^{(i} \nabla_\b^{j} \nabla_\g^{k} \nabla_\d^{l)} \ .
\ee
Here $U_{ijkl}$ is primary and of dimension $-6$. 
One can check that $U_{ijkl}$ is defined up to the gauge transformations
\be \d U_{ijkl} = \nabla_\a^m \xi^\a{}_{ijklm} \ , \label{UGTofXi}
\ee
where $\xi^\a{}_{ijklm} = \xi^\a{}_{(ijklm)}$.

We wish to work out the supercurrent conservation equation in the presence of the compensating $\cO^{*}(4)$ multiplet. 
Let us first consider the general variation of the action with respect to the supergravity and 
matter prepotentials
\be \d S = \int \rd^{6|8}z \, E \, \Big( \d H \,J + \d U_{ijkl} \,
{\mathbb L}^{ijkl} \Big) \ , \label{varO(4)actionSupercurrent}
\ee
where $H$ is the superfield prepotential for conformal supergravity and ${\mathbb L}^{ijkl}$ is some 
superfield of dimension 8 required to be an $\cO(4)$ multiplet as a result of the gauge transformation law \eqref{UGTofXi}. 
As in the previous subsection, in the variation \eqref{varO(4)actionSupercurrent} we have assumed any additional matter fields satisfy their equations of motion. 
The conformal supergravity prepotential transforms under the supergravity gauge transformations 
as eq. \eqref{SUGRAtrans} 
and $\d U_{ijkl}$ should be expressed in terms of three spinor derivatives hitting $\L_{\a ijk}$ 
on dimensional grounds. 
In any case, this should lead to a trace superfield $A^{\a ijk}$ linear in both the 
$\cO{}^{*}(4)$ multiplet $T$ and the $\cO(4)$ multiplet ${\mathbb L}^{ijkl}$.

We expect that $A^\a{}^{ijk}$, 
reduces to the construction in \cite{MR03} after fixing $\Tmult$ to a constant 
and reducing to flat superspace. Taking this into account and 
considering all possible terms linear in $\Tmult$ and ${\mathbb L}^{ijkl}$ 
in conformal superspace, one can construct the most general ansatz for the trace superfield. Then 
demanding the consistency condition \eqref{anomalyConsistEq} fixes it as
\bea A^{\a ijk} &=& \ri \, \Tmult \, \nabla^{\a\b} \nabla_{\b l} {\mathbb L}^{ijkl}
+ \frac{3}{16} \eps^{\a\b\g\d} (\nabla_\b^{(i} \Tmult) \nabla_{\g l} \nabla_{\d p} {\mathbb L}^{jk)lp}
+ \frac{15 \ri}{4} \, (\nabla_{\b l} \Tmult) \nabla^{\a\b} {\mathbb L}^{ijkl} \non\\
&&+ \frac{3}{4} \eps^{\a\b\g\d} (\nabla_\g^{(i} \nabla_{\d l} \Tmult) \nabla_{\b p} {\mathbb L}^{jk)lp} 
+ \frac{5 \ri}{2} (\nabla^{\a\b} \nabla_{\b l} \Tmult) {\mathbb L}^{ijkl}
+ \frac{5}{8} \eps^{\a\b\g\d} (\nabla_\b^{(i} \nabla_{\g l} \nabla_{\d p} \Tmult) {\mathbb L}^{jk)lp} \non\\
&&- 25 X^\a_l \, \Tmult \, {\mathbb L}^{ijkl}
+ 4 \ri W^{\a\b} \, \Tmult \, \nabla_{\b l} {\mathbb L}^{ijkl}
+ 15 \ri \, W^{\a\b} \, (\nabla_{\b l} \Tmult) \, {\mathbb L}^{ijkl} \label{O(4)anomalySuperfieldSUGRA}
\ .
\eea
One can check that this is also primary. 
If we gauge fix $T=1$ and reduce to 
Minkowski superspace,
$\nabla_A \rightarrow (\partial_a , D_\a^i)$ and $W^{\a\b} \rightarrow 0$, we obviously recover \eqref{MRconstruction}.
 
The existence of conserved SU(2) and supersymmetry currents 
is then guaranteed by the results \eqref{O(4)constructionCurrents}.

%%%%%%%%%%%%%%%%%%%%%%%%%%%%%%%%%%%%%%%%%%%%%%%%%%%%%%

\subsubsection{An example: The relaxed hypermultiplet}

It is illustrative to provide an example of a non-conformal supercurrent in curved superspace with  
$\Tmult$ chosen as a compensating superfield. 

In the case of 4D $\cN=2$ Poincar\'e supersymmetry,
the relaxed hypermultiplet \cite{HST83-2} was the first off-shell formulation 
without intrinsic central charge for the massless hypermultiplet. 
This formulation was generalised to 6D $\cN=(1,0)$ supersymmetry 
 in \cite{HST83}. In both cases, 
 the relaxed hypermultiplet was described only in Minkowski superspace. 
To the best of our knowledge, its coupling to supergravity has never been constructed. 
Such a coupling will be given below.
In conformal supergravity one must introduce a compensating $\cO{}^{*}(4)$ multiplet 
as we will show.

To begin with, the relaxed hypermultiplet is described by the superfields 
$L^{ij}$, $L^{ijkl}$ and $\tilde{T}$, subject to the following off-shell constraints
\bsubeq
\bea
\nabla_\a^{(i} L^{jk)} &=& \Tmult \, \nabla_{\a l} {L}^{ijkl} 
+ 5 (\nabla_{\a l} \Tmult) \, {L}^{ijkl} 
\ , \\
\nabla_\a^{(i} L^{jklp)} &=& 0 \ , \\
\nabla_{(\a}^j \nabla^{\phantom{i)}}_{\b) j} \tilde{\Tmult} &=& 0 \ .
\eea
\esubeq
The independent off-shell component fields of the relaxed hypermultiplet can be 
extracted from the above constraints.

The action for the relaxed hypermultiplet may be described in a covariant way using  
the primary superform action \cite{ALR14, BKNT}, which is built out of a 
primary superfield $A_\a{}^{ijk}.$\footnote{The superfield $A_\a{}^{ijk}$ should not 
be confused with $A^{\a ijk}$.} It satisfies the differential constraint
\be \nabla_{(\a}^{(i} A^{\phantom{i)}}_{\b)}{}^{jkl)} = 0 \ .
\ee
One only needs to allow the superfield $A_\a{}^{ijk}$, taking on the role of a Langrangian, to be 
composed of the fields of the relaxed hypermultiplet and the compensating superfield $\Tmult$.
The two supersymmetry invariants 
making up the action for the relaxed hypermultiplet are: (i) 
$I_1$ described by
\be A_\a{}^{ijk} = \Tmult \nabla_{\a l} {\mathbb H}^{ijkl} + 5 (\nabla_{\a l} \Tmult) {\mathbb H}^{ijkl} \ ,
\ee
where
\be {\mathbb H}^{ijkl} = \frac{2}{5} L^{(ij} L^{kl)} 
- \frac{4}{3} \Tmult \, L_p{}^{(i} L^{jkl)p} 
- \frac{15}{7} \Tmult^2 \, L^{mn (ij} L^{kl)}{}_{mn} 
\ ;
\ee
and (ii) $I_2$ described by
\be  A_\a{}^{ijk} = \tilde{\Tmult} \, \nabla_{\a l} {L}^{ijkl} 
+ 5 (\nabla_{\a l} \tilde{\Tmult}) \, {L}^{ijkl} \ . 
\ee
Their linear combination gives the relaxed hypermultiplet action.\footnote{Note that one could 
also choose 
$\Tmult = G^{-1}$ since $G^{-1}$ satisfies the appropriate differential constraint.}

The superspace equations of motion for the relaxed hypermultiplet action are
\be \Tmult \, \nabla_{\a j} L^{ij} + 3 (\nabla_{\a j} \Tmult) L^{ij} 
= \l \nabla_\a^i \Big( \frac{\tilde{\Tmult}}{\Tmult} \Big) \ , \quad {L}^{ijkl} = 0 \ , \label{relaxedHyperEoM}
\ee
where $\l$ is some non-zero 
constant related to the relative coefficients of the invariants. 
The equations of motion \eqref{relaxedHyperEoM} are constructed such that they 
are primary and that they reduce to those 
given in Minkowski superspace in \cite{HST83-2} when $T$ is set to a constant. 
Note that, up to a constant, the equations of motion completely 
determine $\tilde{T}$ in terms of other fields.
The supercurrent is 
\be J = \Tmult L^{ij} L_{ij} 
\ ,
\ee
which is the unique primary scalar that is linear in $T$, quadratic in $L^{ij}$ 
and of dimension 4.

One may verify that the supercurrent conservation equation \eqref{consEqAnomalous} 
recovers \eqref{O(4)anomalySuperfieldSUGRA} with  
a composite multiplet
\be {\mathbb L}^{ijkl} = - \frac{4}{5} L^{(ij} L^{kl)} 
\ ,
\ee
which is an $\cO(4)$ multiplet once one imposes the 
equations of motion for the relaxed hypermultiplet. 
This provides an example verifying the 
supercurrent conservation equation for a compensating $\cO{}^{*}(4)$ multiplet.

%%%%%%%%%%%%%%%%%%%%%%%%%%%%%%%%%%%%%%%%%%%%%%%%%%%%%%

\subsection{Further generalisations}

So far we have found two solutions for the trace superfield $A^{\a ijk}$ which lead to a conserved 
supercurrent and energy-momentum tensor, 
cf. subsections \ref{anomSuperfieldO(2)} and \ref{anomSuperfieldO(4)2.3}. 
One involves an $\cO(2)$ multiplet with a vector multiplet, 
while the other involves an $\cO(4)$ multiplet with an $\cO{}^{*}(4)$ multiplet. It turns out 
there is in an infinite family of solutions that involve the product of an $\cO(n)$ multiplet with 
an $\cO{}^{*}(n)$ multiplet for $n \geq 2$.\footnote{The $\cO{}^{*}(2)$ multiplet is defined 
to be a vector multiplet \cite{KNT17}.} The $\cO{}^{*}(n)$ multiplets were introduced in \cite{KNT17} 
as `dual' to the $\cO(n)$ multiplets in the sense that there exists an action formula that schematically 
involves the product of the two. 
We will describe the defining constraints of these multiplets below 
and introduce the infinite family of non-conformal supercurrents.

The $\cO(n)$ multiplet for $n \geq 1$ is given by a primary superfield $L^{i_1 \cdots i_n} 
$ of dimension $2 n$ satisfying the differential constraint
\be \nabla_\a^{(i_1} L^{i_2 \cdots i_{n+1})}_{\phantom{\a}} = 0 \ . \label{O(n)DefiningConstraint}
\ee
They are off-shell for $n \geq 2$.

The $\cO{}^{*}(3)$ multiplet is described by a 
primary superfield $\Tmult_\a$ of dimension $-3/2$ with the differential constraint
\be \nabla_{(\a}^i \Tmult_{\b)}^{\phantom{j}} = 0 \ ,
\ee
while the $\cO{}^{*}(n)$ multiplet with $n > 4$ is described by a superfield $\Tmult_{i_1 \cdots i_{n-4}}$ 
of dimension $4-2n$ 
satisfying the 
constraint
\be \nabla_\a^j \Tmult_{i_1 \cdots i_{n-5} j}^{\phantom{j}} = 0 \quad \implies 
\quad 
\nabla_{(\a}^j \nabla_{\b) j}^{\phantom{j}} \Tmult_{i_1 \cdots i_{n-4}} = 0 \ . \label{transverseConst}
\ee
The prepotential formulations for these multiplets appeared in \cite{KNT17}.

One can build a primary superfield $A^{\a ijk}$ satisfying \eqref{anomalyConsistEq}
out of an $\cO(3)$ multiplet and an $\cO{}^{*}(3)$ multiplet as follows
\begin{align} A^{\a ijk} &= 
\Tmult_\b \nabla^{\a\b} {\mathbb L}^{ijk}
- \frac{3 \ri}{16} \, \eps^{\ab\g\d} (\nabla_\b^{(i} \Tmult_\g) \nabla_{\d l} {\mathbb L}^{jk) l}
- \frac{\ri}{4} \eps^{\a\b\g\d} (\nabla_\b^{(i} \nabla_{\g l} \Tmult_\d) L^{jk)l}
+ (\nabla^{\a\b} \Tmult_\b) {\mathbb L}^{ijk}
 \non\\
&\qquad \qquad + 6 \, W^{\a\b} \Tmult_\b {\mathbb L}^{ijk} \ .
\end{align}
One can do the same with an $\cO(4+p)$ and an $\cO{}^*(4+p)$ multiplet with $p > 0$ as follows
\bea\label{genAsupFieldO(n)}
A^{\a ijk} &=&
\frac{p}{2(p+4)} \Tmult_{i_1 \cdots i_{p-1}}{}^{(i} \nabla^\a{}_{lmn} {\mathbb L}^{jk) i_1 \cdots i_{p-1} lmn}
+ \ri \, \Tmult_{i_1 \cdots i_p} \nabla^{\a\b} \nabla_{\b l} {\mathbb L}^{ijkl i_1 \cdots i_p} \non\\
&&
+ \frac{p+3}{2(p+4)} (\nabla_\b^{(i} \Tmult_{i_1 \cdots i_p}) \nabla^{\a\b}_{lm} {\mathbb L}^{jk)lmi_1 \cdots i_p}
+ \frac{\ri (p+3)(p+5)}{(p+1)(p+4)} \nabla_{\b i_1} \Tmult_{i_2 \cdots i_{p+1}} \nabla^{\a\b} {\mathbb L}^{ijki_1 \cdots i_{p+1}} \non\\
&&
+ \frac{p+3}{4(p+1)} \eps^{\a\b\g\d} (\nabla_\g^{(i} \nabla_{\d i_1} \Tmult_{i_2 \cdots i_{p+1}}) \nabla_{\b l} {\mathbb L}^{jk)l i_1 \cdots i_{p+1}} \non\\
&&
+ \frac{(p+3) (p+5)}{12 (p+2)(p+1)} \eps^{\a\b\g\d} (\nabla_\b^{(i} \nabla_{\g i_1} \nabla_{\d i_2} \Tmult_{i_3 \cdots i_{p+2}}) {\mathbb L}^{jk) i_1 \cdots i_{p+2}}
\non\\
&& + \frac{\ri (p+3)(p+5)}{3 (p+1)(p+2)} (\nabla^{\a\b} \nabla_{\b i_1} \Tmult_{i_2 \cdots i_{p+1}}) {\mathbb L}^{ijk i_1 \cdots i_{p+1}}
\non\\
&&- \frac{10 (p+1) (p+5)}{p+2} X^\a_{i_1} \Tmult_{i_2 \cdots i_{p+1}} {\mathbb L}^{ijk i_1 \cdots i_{p+1}}
+ 4 \ri \, W^{\a\b} \Tmult_{i_1 \cdots i_p} \nabla_{\b l} {\mathbb L}^{ijkl i_1 \cdots i_p} \non\\
&&
+ \frac{2 \ri (p+3) (p+5)}{(p+1)(p+2)} W^{\a\b} (\nabla_{\b i_1} \X_{i_2 \cdots i_{p+1}}) {\mathbb L}^{ijk i_1 \cdots i_{p+1}} \ ,
\eea
where we have introduced the definition
\be \nabla^{\a\b ij} := \hf \eps^{\a\b\g\d} \nabla_\g^{(i} \nabla_\d^{j)} \ .
\ee
One can check that when $p = 0$, the non-conformal supercurrent corresponding 
to \eqref{genAsupFieldO(n)} agrees with \eqref{O(4)anomalySuperfieldSUGRA}. 
The above general form for the trace superfield corresponds to compensating $\cO^{*}(n)$ 
multiplets. However, for $n \geq 4$ the general form also makes sense if one takes the conformal 
compensator to be an $\cO(n)$ multiplet. 
It can be checked that in either case, after freezing the compensator to a constant and 
reducing to flat superspace, one obtains a conserved $\rm SU(2)$ and, therefore, also a conserved 
supersymmetry current and energy-momentum tensor.
Higher-derivative actions for both cases were described  
in \cite{KNT17}.

%%%%%%%%%%%%%%%%%%%%%%%%%%%%%%%%%%%%%%%%%%%%%%%%%%%%%%
%%%%%%%%%%%%%%%%%%%%%%%%%%%%%%%%%%%%%%%%%%%%%%%%%%%%%%

\section{The supercurrent associated with the dilaton-Weyl multiplet} \label{dilWeylmult}

As mentioned earlier, the tensor multiplet may be used as a conformal 
compensator in supergravity \cite{BSVanP}. 
In conformal superspace, the tensor multiplet is described by a primary superfield $\Phi$ 
of dimension 2 satisfying the following differential constraint
\be \nabla_\a^{(i} \nabla_{\b}^{j)} \Phi = 0 \ . \label{tensorMultiConstraint}
\ee
However, the multiplet is on-shell in the flat case in the sense 
that the constraint \eqref{tensorMultiConstraint} implies $\Box \Phi \equiv \partial^a \partial_a \Phi = 0$ 
and there is no description in terms of an unconstrained superfield for such a multiplet. Despite this, 
it is still possible 
to work out a candidate for the trace superfield $A^{\a ijk}$ for a supergravity theory involving a compensating 
tensor multiplet.\footnote{The fact that it exists is related to the fact that
there is an invariant which is essentially a product of the tensor multiplet and a gauge three-form 
multiplet \cite{BKNT}.} We present this candidate below.

We first observe that we can construct an appropriate primary field $A^{\a ijk}$ as follows
\be \label{tensorAnomalyA}
A^{\a ijk} = \frac{\ri}{3} \Phi \nabla_\b^{(i} {\mathbb H}^{\a\b jk)}
+ \ri (\nabla_\b^{(i} \Phi) {\mathbb H}^{\a\b jk)} \ ,
\ee
where ${\mathbb H}^{\a\b ij} = {\mathbb H}^{[\a\b] (ij)}$ is a primary superfield of dimension 3. Now we need to 
impose additional constraints on ${\mathbb H}^{\a\b ij}$ in order for the trace superfield $A^{\a ijk}$ to  
both satisfy the consistency condition \eqref{anomalyConsistEq} and imply the existence of a conserved supersymmetry current.
One can show that the consistency condition \eqref{anomalyConsistEq} is satisfied if we impose the constraint
\be \nabla_\a^{(i} {\mathbb H}^{\b\g jk)} = - \frac{2}{3} \d_\a^{[\b} \nabla_\d^{(i} {\mathbb H}^{\g]\d jk)} \ . \label{HconstA}
\ee
One can check that eq. \eqref{HconstA} is a primary constraint.

There exists another primary constraint that one can impose on ${\mathbb H}^{\a\b ij}$ and it is
\be \nabla_\a^{(i} \nabla_{\b k} {\mathbb H}^{\a\b j) k} + 3 \ri \nabla_{\a\b} {\mathbb H}^{\a\b ij} = 0 \ . \label{HconstB}
\ee
The constraints \eqref{HconstA} and \eqref{HconstB} are exactly the primary constraints that ensure  ${\mathbb H}^{\a\b ij}$ 
describes the lowest dimension component of a closed four-form \cite{ALR14}.

One can check that in the flat case
with the tensor multiplet set to unity, i.e.  $\Phi = 1$, and using the constraint \eqref{HconstB}, 
the descendent $A^{ij}$, defined by \eqref{Descendents2.6a}, is
\be A^{ij} = \frac{3}{16} D_{\a k} A^{\a ijk} = \frac{1}{4} \partial_{\a\b} {\mathbb H}^{\a\b ij} \ .
\ee
Since $A^{ij}$ is a divergence we have a conserved 
$\rm SU(2)$ current, together with a conserved supersymmetry current and energy-momentum 
tensor according to the analysis of 
subsection \ref{MinkAnalysSection}. These are
\bsubeq
\begin{align}
j_{\a\b}{}^{ij} &= V_{\a\b}{}^{ij} - \ri {\mathbb H}_{\a\b}{}^{ij} \ , \quad \partial^{\a\b}  j_{\a\b}{}^{ij} = 0 \ , \\
S_{\a\b ,}{}_\g^i &= \S_\g^i{}_{,\a\b} - \frac{2 \ri}{3} D_{\g j} {\mathbb H}_{\a\b}{}^{ij} \ , \quad \partial^{\a\b} S_{\a\b ,}{}_\g^i = 0 \ , \\
T_{\a\b , \g\d} &= \cT_{\a\b , \g\d} - \frac{\ri}{6} D_{\a i} D_{\b j} {\mathbb H}_{\g\d}{}^{ij}
- \frac{\ri}{6} D_{\g i} D_{\d j} {\mathbb H}_{\a\b}{}^{ij} \ , \quad \partial^{\a\b}  T_{\a\b , \g\d} = 0 \ .
 \end{align}
 \esubeq
One should note that the supersymmetry current is not gamma-traceless and neither is the energy-momentum tensor 
traceless. To prove conservation of the energy-momentum tensor one uses
\be \partial^{\a\b} D_{\a i} D_{\b j} {\mathbb H}_{\g\d}{}^{ij} = \partial^{\a\b} D_{\g i} D_{\d j} {\mathbb H}_{\a\b}{}^{ij} \ ,
\ee
which follows from the differential constraints on ${\mathbb H}_{\a\b}{}^{ij}$, eqs. \eqref{HconstA} and \eqref{HconstB}.

Remarkably, if we deform the constraint defining the tensor multiplet to
\be \nabla_\a^{(i} \nabla_\b^{j)} \Phi = \ri \, {\mathbb H}_{\a\b}{}^{ij} \ ,
\ee
the postulated superfield \eqref{tensorAnomalyA} still satisfies the consistency 
condition \eqref{anomalyConsistEq} but a conserved supersymmetry current is no longer guaranteed.

%%%%%%%%%%%%%%%%%%%%%%%%%%%%%%%%%%%%%%%%%%%%%%%%%%%%%%

\subsection{An example: Non-abelian gauge theory involving a compensating tensor multiplet}

For an illustrative example of a non-conformal supercurrent in curved superspace, 
we consider non-abelian gauge theory involving a compensating tensor multiplet. 
We refer the reader to appendix \ref{YMmultiplet} for details on 
the description of the Yang-Mills multiplet in conformal superspace. 
The action for the theory is composed of two parts: (i) a higher-derivative 
non-abelian vector multiplet action; 
and (ii) the Yang-Mills action which involves the tensor multiplet and contains the term 
$\s \Tr({\bm f}^{ab} {\bm f}_{ab})$ at the component level \cite{BSVanP}.  
Here $\s$ is the component projection of $\Phi$ 
and ${\bm f}_{ab}$ is the field strength of the non-abelian gauge field.

The higher-derivative non-abelian vector multiplet action is the 
non-abelian extension of the supersymmetric $F \Box F$ action 
mentioned in subsection \ref{anomSuperfieldO(2)}. It was  
described in Minkowski superspace in \cite{ISZ05} and in conformal 
superspace in \cite{BKNT}. The supercurrent is
\be
J = \frac{3}{8} \Tr\Big( X^{ij} X_{ij} + \frac{4\ri}{3} W^{\a i} \bm \nabla_{\a\b} W^{\b}_i 
+ \frac{2}{3} F_\a{}^\b F_\b{}^\a \Big) \ . \label{supercurrentFBoxF}
\ee
It is the unique dimension four primary superfield quadratic in the fields of the vector multiplet.  
The equation of motion for the vector multiplet is
\be \bm \cG^{ij} =0 ~,\label{2.54}
\ee
where we have introduced the superfield
\bea
\bm \cG^{ij} := 
 \bm \nabla^a \bm \nabla_a
X^{ij} 
&-& 2 \ri [W^{\a (i} , \bm \nabla_{\a\b} W^{\b j)}]
- \frac{3}{2} [X^{k(i} , X^{j)}{}_k ] 
- 2 Y_\a{}^\b{}^{ij} F_\b{}^\a  \non \\
&+& \frac{5}{2} X^{\a (i} \overleftrightarrow{\bm \nabla}_{\a\b} W^{\b j)} \ ,
\eea
which is constructed to be primary of dimension 4 and to satisfy 
${\bm \nabla}_\a^{(i} {\bm \cG}^{jk)} = 0$.\footnote{JN is grateful to Daniel Butter for checking this result 
using the computer algebra program {\it Cadabra}.}
It can be checked that the supercurrent \eqref{supercurrentFBoxF} is conserved,
\be
\nabla^{\a ijk} J = \ri \Tr \Big( W^{\a (i} \bm \cG^{jk)} \Big) = 0 \ ,
\ee
as a consequence of the equation of motion \eqref{2.54}.

We can now check the non-conformal conservation condition if we consider the higher-derivative 
non-abelian vector multiplet action coupled to the Yang-Mills action. The Yang-Mills 
action was described in conformal superspace in \cite{BKNT} by making use of a closed six-form with the lowest component 
given by the primary superfield
\be A_\a{}^{ijk} = \eps_{\a\b\g\d} V^{\b (i} H^{\g\d jk)} \ , \quad H^{\a\b ij} := \ri \Tr(W^{\a (i} W^{\b j)}) \ ,
\ee
where $V^{\a i}$ is the {\it constrained}
prepotential for the tensor multiplet \cite{Sokatchev88, LT-M12}, which satisfies\footnote{Invariance under 
gauge transformations of the prepotential was shown in \cite{BKNT}.}
\be \nabla_\a^{(i} V^{\b j)} - \frac{1}{4} \d_\a^\b \nabla_\g^{(i} V^{\g j)} = 0 \ , \quad \Phi = \nabla_{\a i} V^{\a i} \ , 
\quad K^A \Phi = 0\ ,\quad{\mathbb D}\Phi=2\Phi  \ .
\ee
It should be noted that the primary superfield $H^{\a\b ij}$ satisfies the same differential 
constraints as ${\mathbb H}^{\a\b ij}$ does.

For the combined action incorporating the higher-derivative 
non-abelian vector multiplet action and the Yang-Mills action, the equation of motion for the vector multiplet 
becomes
\be \bm \cG^{ij} + \l \Big( \Phi X^{ij} + \ri (\nabla_\a^{(i} \Phi) W^{\a j)} \Big) = 0 \ ,
\ee
where $\l$ is a coupling constant. We also need to know the supercurrent $J$ for the 
combined theory. Interestingly, the supercurrent $J$ does not obtain a contribution 
from the Yang-Mills action. The point is that such a supercurrent 
would have to be linear in $\Phi$ and quadratic in the fields of the vector multiplet and 
no such scalar superfield of dimension 4 exists. Therefore, much like the $BF$ invariant, 
the Yang-Mills action can have no dependence on the supergravity prepotential $H$ and one can use 
the supercurrent of the higher-derivative Yang-Mills action for the combined system.
Now using the supercurrent \eqref{supercurrentFBoxF} and the equation of motion we find
\be \nabla^{\a ijk} J = \frac{\ri}{3} \Phi \nabla_\b^{(i} {\mathbb H}^{\a\b jk)} + \ri (\nabla_\b^{(i} \Phi) {\mathbb H}^{\a\b jk)} \ , 
\quad {\mathbb H}^{\a\b ij} = \l H^{\a\b ij} \ ,
\ee
which verifies the non-conformal supercurrent equation in the presence of a compensating tensor multiplet, 
cf. \eqref{tensorAnomalyA}.

%%%%%%%%%%%%%%%%%%%%%%%%%%%%%%%%%%%%%%%%%%%%%%%%%%%%%%

\subsection{The dilaton-Weyl multiplet}

We now discuss some subtleties about the non-conformal supercurrent just presented. 
As we have seen 
in previous sections, the 
supercurrent may be understood in terms of the variation of an action with respect to the conformal 
supergravity prepotential and possibly the prepotential of some supermultiplet that is to take on the role as a 
compensator. However, we obviously bump into a problem when we choose the compensator to be a tensor multiplet which has no 
prepotential formulation.

The tensor multiplet is quite special because its defining 
constraint \eqref{tensorMultiConstraint} allows one to express the super-Weyl tensor in terms of the fields of the tensor multiplet,
\be
W_{abc} = -\frac{1}{4} H_{abc} - \frac{\ri}{32} (\tilde{\g}_{abc})^{\g\d} \nabla_\g^k \nabla_{\d k} \Phi \ ,
\ee
where $H_{abc}$ is the three-form field strength of the tensor multiplet. One should keep in mind that the combined system,
tensor $+$ Weyl-multiplet, 
is off-shell (with $40+40$ degrees of freedom) and upon replacing the covariant fields of the Weyl multiplet 
with those of the tensor multiplet leads to what is known as the dilaton-Weyl or type II Weyl 
multiplet \cite{BSVanP}. One expects that the dilaton-Weyl multiplet should 
possess a prepotential formulation, albeit potentially taking a different form than that of 
the standard Weyl multiplet. 
We do not derive the details of such 
a formulation here but we wish to emphasise some important points below.

It is instructive to consider a superconformal action 
that may be described in standard conformal supergravity without a tensor multiplet, 
which possesses the supercurrent $J$ with the usual conservation 
condition \eqref{ConformConservEqn}.\footnote{An example is provided 
by the linear (or $\cO(2)$) multiplet action \cite{GL85, BSVanP} 
where the supercurrent is given (up to normalisation) by $J = G$.} 
We can always replace the fields of the standard Weyl multiplet in the action 
with those of the dilaton-Weyl multiplet, which involves the 
tensor multiplet and thus gives a new action.  
However, this should only lead to a rewriting of the conservation condition on the 
supercurrent:
\be
\nabla^{\a ijk} J = \nabla^{\a ijk} (\Phi \frac{J}{\Phi}) = \frac{\ri}{3} \Phi \nabla_\b^{(i} \tilde{H}^{\a\b jk)} 
+ \ri (\nabla_\b^{(i} \Phi) \tilde{H}^{\a\b jk)} = 0 \ , \label{ConsvCondFromJtoH}
\ee
where
\be \tilde{H}_{\a\b}{}^{ij} = \ri \nabla_\a^{(i} \nabla_\b^{j)} \Big( \frac{J}{\Phi} \Big)
\ee
and $\tilde{H}_{\a\b}{}^{ij}$ satisfies the differential constraints \eqref{HconstA} and \eqref{HconstB}. 
We see that for every such theory there always exists a superfield $\tilde{H}^{\a\b ij}$ subject to the 
conservation condition \eqref{ConsvCondFromJtoH}.

The observation that the conservation of the supercurrent $J$ can be rewritten in terms of 
the superfield $\tilde{H}^{\a\b ij}$ is important since the Yang-Mills action 
does not possess a supercurrent $J$ as discussed earlier. One can instead 
understand the superfield 
$H^{\a\b ij} = \ri \Tr(W^{\a (i} W^{\b j)})$ as the supercurrent
in the dilaton-Weyl multiplet. This is consistent with the fact that the action 
is linear in the tensor multiplet and built out of covariant derivatives of $H^{\a\b ij}$. Furthermore, the 
superfield $H^{\a\b ij}$ corresponding to the Yang-Mills action action satisfies the conservation 
condition \eqref{ConsvCondFromJtoH} when the equation of motion for the Yang-Mills multiplet is enforced.

For the reasons mentioned above, one should think of the superfield $J^{\a\b ij}$ satisfying the constraints \eqref{HconstA} 
and \eqref{HconstB} (with ${\mathbb H}^{\a\b ij}$ replaced with $J^{\a\b ij}$) 
and the on-shell conservation condition
\be
\frac{\ri}{3} \Phi \nabla_\b^{(i} J^{\a\b jk)} 
+ \ri (\nabla_\b^{(i} \Phi) J^{\a\b jk)} = 0
\label{3.20}
\ee
as the supercurrent for a theory coupled to the dilaton-Weyl multiplet.

The dilaton-Weyl multiplet is expected to be described by an unconstrained prepotential $h^{\a\b}{}_{ij} $ such 
that its infinitesimal displacement generates
the following variation of an action 
\be \d S = \int \rd^{6|8}z \, E \, \d h^{\a\b}{}_{ij}  \, J_{\a\b}{}^{ij} ~.
\ee
The constraints \eqref{HconstA} and \eqref{HconstB} imposed on the supercurrent $J_{\a\b}{}^{ij} $ 
should be the conditions of the invariance of the action $S$ under 
certain gauge transformations of the gravitation superfield $h^{\a\b}{}_{ij}$. In fact these conditions follow 
from the gauge transformations
\bea
\d h_{\a\b}{}^{ij} = \nabla_{\g k} \L_{\a\b}{}^{\g ijk} + \ri {\nabla}_{[\a}^{(i} {\nabla}_{\b] k} \L^{j) k} - 3 \nabla_{\a\b} \L^{ij} \ ,
\eea
where the gauge parameters are primary and satisfy the conditions
\be \L_{\a\b}{}^{\g ijk} = \L_{[\a\b]}{}^{\g (ijk)} \ , \quad \L_{\a\b}{}^{\b ijk} = 0 \ , \quad \L^{ij} = \L^{(ij)} \ .
\ee
The conservation condition \eqref{3.20} follows from the supergravity gauge transformations
\be \d h_{\a\b}{}^{ij} = \ri \Phi \nabla_{[\a k} \L_{\b]}{}^{ijk} - 2 \ri (\nabla_{[\a k} \Phi) \L_{\b]}{}^{ijk} \ , 
\quad \L_\a{}^{ijk} = \L_\a{}^{(ijk)} \ .
\ee
One can see that in the Minkowski superspace limit, the supercurrent $J^{\a\b ij}$ satisfies the constraints
\be D_\a^{(i} J^{\b\g}{}^{jk)} = 0 \ , \quad \partial^{\a\b} J_{\a\b}{}^{ij} = 0 \ .
\ee
Corresponding to the supercurrent put forward in \cite{HST83,HU87}. One can also check that $J^{\a\b ij}$ 
possesses $40+40$ degrees of freedom.

Suppose a matter action $S$ couples to a compensator, for instance the linear multiplet. Then the conservation equation \eqref{3.20} gets deformed to take the form
\bea
\frac{\ri}{3} \Phi \nabla_\b^{(i} J^{\a\b jk)} 
+ \ri (\nabla_\b^{(i} \Phi) J^{\a\b jk)} = A^{\a ijk} ~.
\eea
The consistency condition \eqref{anomalyConsistEq} follows from the above conservation 
condition keeping in mind the constraints \eqref{HconstA} and \eqref{HconstB} 
imposed on $J^{\a\b ij}$, as well as the constraint \eqref{tensorMultiConstraint} on $\Phi$.
 Using the results of subsection \ref{MinkAnalysSection}, 
we find that in the Minkowski superspace limit with $\Phi = 1$
\be \partial^{\a\b} J_{\a\b}{}^{ij} = - 4 A^{ij} \ , \quad \partial^{\a\b} \hat{\S}_{\a\b ,}{}_{\g}^k = 4 \ri \L_\g^k \ , 
\quad \partial^{\a\b} \hat{T}_{\a\b , \g\d} = 4 \ri \cA_{\g\d} \ ,
\ee
where we have defined
\be 
\hat{\S}_{\a\b ,}{}_{\g}^i := - \frac{2 \ri}{3} D_{\g j} J_{\a\b}{}^{ij} \ , \quad
\hat{T}_{\a\b , \g\d} := - \frac{\ri}{6} D_{\a i} D_{\b j} J_{\g\d}{}^{ij}- \frac{\ri}{6} D_{\g i} D_{\d j} J_{\a\b}{}^{ij} \ .
\ee
This tells us that we still require $\L_\g^k$ and $\cA_{\g\d}$ to be divergences in order for a {\it conserved} 
supersymmetry current and energy-momentum tensor to exist. However, 
now the supersymmetry current contains a 
gamma-trace component and the energy-momentum 
tensor contains a trace in addition to any contribution from the 
trace superfield $A^{\a ijk}$, which can be chosen to be any of the trace 
superfields derived in section \ref{(1,0)AnomSupercurrent}.

%%%%%%%%%%%%%%%%%%%%%%%%%%%%%%%%%%%%%%%%%%%%%%%%%%%%%%
%%%%%%%%%%%%%%%%%%%%%%%%%%%%%%%%%%%%%%%%%%%%%%%%%%%%%%

\section{The $\cN = (2, 0)$ non-conformal supercurrent} \label{(2,0)supercurrent}

In this section, we discuss the $\cN = (2, 0)$ superconformal current and 
put forward an $\cN = (2, 0)$ extension of the $\cN = (1, 0)$ non-conformal supercurrent 
based on a compensating tensor multiplet.

We first review some basic 
notation and conventions in regards to $\cN = (2, 0)$ supersymmetry. 
A symplectic Majorana spinor $\Psi_{\hi }$, decomposed as in \cite{BKNT}, has Weyl components 
that satisfy the reality conditions
\begin{align}\label{eq:SympMaj4c}
\overline{\psi^{\alpha \hi }} = \psi^\alpha_{\hi }~, \qquad
\overline{\chi_{\alpha \hi }} = \chi_\alpha^{\hi }~,
\end{align}
where $\hi  = 1,\dots,4$ are USp(4) indices corresponding to the 
$R$-symmetry group. The $\rm USp(4)$ indices are raised and lowered as 
\begin{align}
\Psi^{\hi } = \Omega^{\hi  \hj} \Psi_{\hj}~, 
\qquad \Psi_{\hi } = \Omega_{\hi  \hj} \Psi^{\hj}~, \quad \Omega_{\hi \hj} \Omega^{\hj\hk} = \d_{\hi }^{\hk} \ ,
\end{align}
where $\Omega^{\hi \hj} = \Omega^{[\hi \hj]}$ is a symplectic metric of USp(4).\footnote{$\cN = (1, 0)$  
is recovered by restricting $i = 1, 2$ and setting $\Omega^{ij} = \eps^{ij}$.} 
It satisfies
\bsubeq
\bea
\eps_{\hi \hj\hk\hl} &=& 3 \Omega_{\hi [\hj} \Omega_{\hk\hl]} \quad \implies 
\quad \Omega^{\hi \hj} = -\hf \eps^{\hi \hj\hk\hl} \Omega_{\hk\hl} \ , \\
\Omega^{[\hi \hj} \Phi^{\hk\hl]} &=& \hf (\Omega^{[\hi \hj} \Phi^{\hk]\hl} 
- \Omega^{\hl[\hk} \Phi^{\hi \hj]}) = 0 \ ,
\eea
\esubeq
where $\Phi^{\hi \hj} = \Phi^{[\hi \hj]}$ is an antisymmetric rank 2 isospinor 
such that $\Phi^{\hi \hj} \Omega_{\hi \hj} = 0$. Note that 
every antisymmetric rank 2 USp(4) tensor $U^{\hi\hj}$ admits the decomposition 
$U^{\hi \hj} = \Phi^{\hi \hj} + \Omega^{\hi \hj} U$. Finally, the chiral $\cN = (2, 0)$ 
supersymmetry algebra is
\be \{ D_\a^{\hi } , D_\b^{\hj} \} = - 2 \ri \Omega^{\hi \hj} \partial_{\a\b} 
\equiv - 2 \ri \Omega^{\hi \hj} (\g^a)_{\a\b} \partial_a \ .
\ee

The $\cN = (2, 0)$ conformal supercurrent is described by the USp(4) tensor superfield 
$J^{\hi\hj , \hk\hl} = J^{[\hi \hj],[\hk\hl]} = J^{\hk\hl, \hi \hj} 
= - 2 J^{\hk[\hi  , \hj] \hl}$, which is 
completely traceless with respect to the sympletic metric $\Omega_{\hi \hj}$ of USp(4), and 
satisfies the superspace conservation condition \cite{HST83}
\be D_\a^{\hm} J^{\hi\hj, \hk\hl} - \Omega^{\hm[\hi} \Psi_\a^{\hj] , \hk\hl} 
- \frac{1}{4} \Omega^{\hi\hj} \Psi_\a^{\hm , \hk\hl} - \Omega^{\hm [\hk} \Psi_\a^{\hl],\hi\hj} - \frac{1}{4} \Omega^{\hk\hl} \Psi_\a^{\hm , \hi\hj} = 0 \ , \label{2,0cscrtCEq}
\ee
where $\Psi_\a^{\hi , \hj\hk} = \Psi_\a^{\hi , [\hj\hk]}$ is completely traceless, $\Psi_\a^{\hi,\hj\hk} \Omega_{jk} = \Psi_\a^{\hi,}{}^{\hj\hk} \Omega_{\hi\hj} = 0$. The condition \eqref{2,0cscrtCEq} is a constraint 
on the completely traceless part of $D_\a^{\hm} J^{\hi\hj , \hk\hl}$ and it 
fixes $\Psi_\a^{\hi}{}^{, \hj\hk} = \frac{4}{7} D_{\a \hl} J^{\hl\hi , \hj \hk}$.

We can now insert a superfield $A_\a^\hm{}^{, \hi\hj}{}^{, \hk\hl}$ in the conservation 
equation \eqref{2,0cscrtCEq} as follows:\footnote{This 
was also considered in \cite{MR03} but with only a partial solution.}
\be  D_\a^\hm J^{\hi\hj, \hk\hl} - \Omega^{\hm[\hi} \Psi_\a^{\hj] , \hk\hl} 
- \frac{1}{4} \Omega^{\hi\hj} \Psi_\a^{\hm , \hk\hl} - \Omega^{\hm [\hk} \Psi_\a^{\hl],\hi\hj} 
- \frac{1}{4} \Omega^{\hk\hl} \Psi_\a^{\hm , \hi\hj} = A_\a^\hm{}^{,\hi\hj}{}^{,\hk\hl} \ ,
\ee
where we require the trace superfield to satisfy the symmetry properties
\be A_\a^\hm{}^{,\hi\hj}{}^{,\hk\hl} = A_\a^\hm{}^{,[\hi\hj]}{}^{,[\hk\hl]} = A_\a^\hm{}^{,\hk\hl}{}^{,\hi\hj} \ , 
\quad A_\a^\hm{}^{,\hi\hj}{}^{,\hk\hl} \Omega_{\hi\hj} = A_\a^\hm{}^{,[\hi\hj}{}^{,\hk\hl]} = 
A_\a^{[\hm}{}^{,\hi\hj]}{}^{,\hk\hl} = 0 \ , 
\ee
and the integrability condition\footnote{This condition follows from requiring 
closure of supersymmetry on $J^{\hi\hj, \hk\hl}$.}
\be
D_{\a}^{\hm} A_{\b}^{\hn}{}^{, \hi\hj}{}^{, \hk\hl} + D_{\b}^{\hn} A_{\a}^{\hm}{}^{, \hi\hj}{}^{, \hk\hl} - ({\rm traces}) = 0 \ , 
\label{intCondA}
\ee
where $({\rm traces})$ represents all terms proportional to the metric $\Omega^{\hi\hj}$ consistent with the symmetry properties 
of $A_\a^\hm{}^{, \hi\hj}{}^{, \hk\hl}$.

We now put forward a candidate for the superfield $A_\a^\hm{}^{, \hi\hj}{}^{, \hk\hl}$ that is 
analogous to the non-conformal supercurrent based on a compensating $\cN = (1,0)$ 
tensor multiplet. As a compensator, we will 
choose  an $\cN = (2, 0)$ tensor multiplet, 
which is described by an antisymmetric and $\Omega$-traceless 
superfield $\Phi^{\hi\hj}$, 
\be
\Phi^{(\hi\hj)} = \Phi^{\hi\hj} \Omega_{\hi\hj} = 0 \ ,
\ee
satisfying the following differential constraint
\be D_\a^{\hi} \Phi^{\hj\hk} - \Omega^{\hi[\hj} \l_\a^{\hk]} - \frac{1}{4} \Omega^{\hj\hk} \l_\a^{\hi} = 0 \ .\label{(2,0)TensorMultdefConst}
\ee
The constraint \eqref{(2,0)TensorMultdefConst} 
eliminates the completely traceless part of $D_\a^{\hi} \Phi^{\hj\hk}$ and 
determines $\l_\a^\hi = \frac{4}{5} D_{\a \hj} \Phi^{\hj\hi}$. 

Inspired by the $\cN = (1, 0)$ case, we write down the following candidate for the 
trace superfield
\be A_\a^{\hm}{}^{,\hi\hj}{}^{,\hk\hl} = \Phi^{\hi\hj} H_\a^{\hm , \hk\hl} + \Phi^{\hk\hl} H_\a^{\hm , \hi\hj} - ({\rm traces}) \ , \label{PhiAnomProposal}
\ee
where $H_\a^{\hi}{}^{, \hj\hk} = H_\a^{\hi}{}^{, [\hj\hk]}$ is completely traceless and satisfies the constraint
\be D_{\a}^{\hi }H_{\b}^{\hj}{}^{, \hk\hl} + D_{\b}^{\hj }H_{\a}^{\hi}{}^{, \hk\hl} - ({\rm traces}) = 0 
\ .
\ee
The above constraint ensures that $A_\a^{\hm}{}^{,\hi\hj}{}^{,\hk\hl}$ satisfies its integrability condition \eqref{intCondA}. It is 
important to mention that the 
superfield $H_\a^{\hi}{}^{, \hj\hk}$ is still very long and should be constrained to ensure the existence of a 
conserved supersymmetry current. In analogy with the $\cN = (1, 0)$ case, we can constrain 
$H_\a^{\hi}{}^{, \hj\hk}$ to be the lowest component of a closed four-form.
This ensures that there exists an $\cN = (1, 0)$ component field 
which is the lowest component of a four-form in the $\cN = (1, 0)$ case. The postulated trace superfield
\eqref{PhiAnomProposal} is expected to be the $\cN = (2, 0)$ extension of the one in 
the $\cN = (1, 0)$ case. The additional constraints 
on $H_\a^{\hi}{}^{, \hj\hk}$ and the closed superform are described in appendix \ref{c4Form}.

%%%%%%%%%%%%%%%%%%%%%%%%%%%%%%%%%%%%%%%%%%%%%%%%%%%%%%
%%%%%%%%%%%%%%%%%%%%%%%%%%%%%%%%%%%%%%%%%%%%%%%%%%%%%%

\section{Discussion} \label{discussion}

In this paper, we have presented various non-conformal supercurrents by finding 
solutions to the deformed conservation equation \eqref{consEqAnomalous}. Remarkably, 
we have managed to uncover an infinite number of solutions that are based on $\cO(n)$ 
multiplets. 
The $n = 2$ case corresponds to choosing 
the well-known linear multiplet as a compensator. 
For $n > 2$ the possible compensators have not been extensively 
considered in detail before. 
Nevertheless, their usefulness was demonstrated in the description of the relaxed 
hypermultiplet given in this paper and such compensators can be 
used in the description of higher derivative actions (see 
the discussion  
section of \cite{KNT17}).
In this light, it would be interesting if our results could 
be used to derive the equations of motion for higher derivative actions. Furthermore, 
the results in this paper should have analogues in lower dimensions and it would 
be interesting to work out their details in future work.

We explored the curious case of using the tensor multiplet as a compensator 
in section \ref{dilWeylmult}. Since coupling the Weyl multiplet to a tensor multiplet leads to a variant 
version of the Weyl multiplet, called the dilaton-Weyl multiplet, the supercurrent 
for the combined system needed to be modified. As an application of this 
supercurrent, we give the superspace equations of motion for minimal Poincar\'e 
supergravity \cite{CvP11} below.

Minimal Poincar\'e supergravity is derived from 
the action for the linear (or $\cO(2)$) multiplet in conformal supergravity \cite{BSVanP}, which 
in superspace can be described by the full superspace integral
\be S_{\rm L} := \int \rd^6 x \, \rd^8 \q \, E \, \r_{\a i} {\mathbb W}^{\a i} \ .
\ee
Here $\r_{\a i}$ is the prepotential for the linear multiplet and ${\mathbb W}^{\a i}$ 
corresponds to an off-shell vector multiplet built out of the fields of the 
linear multiplet as follows \cite{BNT-M}
\bea {\mathbb W}^{\a i} &=& \frac{1}{G} \nabla^{\a\b} \U_\b^i
+ \frac{4}{G} (W^{\a\b} \U_\b^i + 10 \ri X^\a_j G^{ij})
- \frac{1}{2 G^3} G_{jk} (\nabla^{\a\b} G^{ij}) \U_\b^k \non\\
&&+ \frac{1}{2 G^3} G^{ij} F^{\a\b} \U_{\b j}
+\frac{\ri}{16 G^5} \U_{\b j} \U_{\g k} \U_{\d l} G^{ij} G^{kl} \ ,
\eea
where we have defined $\U_\a^i := \frac{2}{3} \nabla_{\a j} G^{ij}$ 
and $F_{\a\b} := \frac{\ri}{4} \nabla_{[\a}^k \U^{\phantom{k}}_{\b] k}$. The equations 
of motion in the standard Weyl multiplet read\footnote{The first equation can, in principle, 
be derived by varying the action with respect to $\rho_{\alpha i}$; this 
is tedious because of the explicit $\rho$-dependence of ${\mathbb W}^{\a i}$. 
Alternatively, one can construct the most general primary field with the same index 
structure and weight ($3/2$) as ${\mathbb W}^{\alpha i}$. 
The equation of motion for the supergravity multiplet is, 
${\delta S\over\delta H}\equiv J=0$. The supercurrent $J$ is a covariant 
expression built from the linear multiplet. Dimensional arguments fix it to be proportional to $G$. }
\be {\mathbb W}^{\a i} = 0 \ , \quad J \propto G = 0 \ .
\ee
These equations of motion are obviously incompatible with the 
requirement that the linear multiplet is a conformal compensator since $G$ 
needs to be set to a non-vanishing constant. To remedy this, we should instead 
replace the Weyl multiplet with the dilaton-Weyl multiplet. Upon doing so, the 
equations of motion become
\be {\mathbb W}^{\a i} = 0 \ , \quad 
{\mathbb H}_{\a\b}{}^{ij} \propto \nabla_\a^{(i} \nabla_\b^{j)} \Big(\frac{G}{\Phi}\Big) = 0 \ ,
\ee
which are now consistent. The 
superspace equations of motion for gauged minimal supergravity \cite{BCSvP12}
can be written down by considering a linear combination of the linear multiplet, 
the $BF$ and the Yang-Mills multiplet actions, and using the results in this paper. 
They are 
\bsubeq
\bea 
{\mathbb W}^{\a i} &=& - 2 g \, W^{\a i} \ , \\
\Phi X^{ij} + \ri (\nabla_\a^{(i} \Phi) W^{\a j)}  &=& - g \, G^{ij} \ , \\
\frac{1}{4} \, \eps^{\a\b\g\d} \nabla_\g^{(i} \nabla_\d^{j)} \Big(\frac{G}{\Phi}\Big) 
&=& W^{\a (i} W^{\b j)} \ ,
\eea
\esubeq
where $W^{\a i}$ describes an abelian vector multiplet, 
and $g$ is a coupling constant.

It is worth discussing the results in this paper in the context of Weyl anomalies. 
When one lifts a classical conformal field theory to curved space the resulting theory 
remains independent of any compensating scalar field. However, the conformal symmetry 
is anomalous at the quantum level. In the Weyl invariant formulation for gravity, 
the presence of conformal (or Weyl) anomalies is equivalent to the fact that the 
effective action acquires dependence on some compensator. The situation with 
supersymmetric field theories is analogous. Given a superconformal field theory, its 
action is independent of any compensator. The presence of superconformal (or super-Weyl) 
anomalies is equivalent to the fact that the effective action acquires dependence 
on a {\it special} compensator. In the case of 4D $\cN=1$ superconformal theories, it 
was argued in \cite{GGS} that the chiral scalar compensator of old minimal supergravity 
couples to the super-Weyl anomalies. The 4D $\cN=1$ super-Weyl anomalies
were studied in \cite{BPT,BK86}.  For 4D $\cN=2$ superconformal theories, 
the super-Weyl anomalies are associated with the vector multiplet compensator
\cite{K13}.

It is natural to ask if any of the non-conformal supercurrents 
correspond to those associated with the super-Weyl anomalies in six dimensions. 
Here it is important to realise that unlike in four dimensions, the super-Weyl anomalies 
in six dimensions should be accompanied with Lorentz anomalies. However, 
each of the non-conformal supercurrents in this paper describe a conserved 
energy-momentum tensor with no Lorentz anomaly. This includes the non-conformal 
supercurrent corresponding to the conservation equation \eqref{1.4} and 
given in \cite{MR03}.\footnote{In this sense it is very much like its counterpart described by 
\eqref{1.4}, which does not couple to the 4D $\cN=2$ super-Weyl anomalies \cite{K13}.} 
The absence of a Lorentz anomaly is evident from the fact that we assumed the supergravity 
actions were invariant under supergravity gauge transformations. Therefore, we need to 
change our set-up and this will be discussed elsewhere.

%%%%%%%%%%%%%%%%%%%%%%%%%%%%%%%%%%%%%%%%%%%%%%%%%%%%%%
%%%%%%%%%%%%%%%%%%%%%%%%%%%%%%%%%%%%%%%%%%%%%%%%%%%%%%
%%%%%%%%%%%%%%%%%%%%%%%%%%%%%%%%%%%%%%%%%%%%%%%%%%%%%%
$~$\\
\noindent
{\bf Acknowledgements:}\\
 JN is grateful to Daniel Butter for 
useful discussions. 
 SMK and JN thank the Galileo Galilei Institute for Theoretical Physics for hospitality 
and the INFN for partial support  in September 2016. SMK acknowledges 
the Albert Einstein Institute for warm hospitality during part of this work. 
The work of SMK is supported in part by the Australian Research Council, 
project No.\,DP160103633.
JN is supported by a Humboldt research fellowship of 
the Alexander von Humboldt Foundation.
JN and ST acknowledge support from GIF -- the German-Israeli Foundation 
for Scientific Research and Development.

%%%%%%%%%%%%%%%%%%%%%%%%%%%%%%%%%%%%%%%%%%%%%%%%%%%%%%
%%%%%%%%%%%%%%%%%%%%%%%%%%%%%%%%%%%%%%%%%%%%%%%%%%%%%%

\appendix

%%%%%%%%%%%%%%%%%%%%%%%%%%%%%%%%%%%%%%%%%%%%%%%%%%
%%%%%%%%%%%%%%%%%%%%%%%%%%%%%%%%%%%%%%%%%%%%%%%%%%

\section{The geometry of $\cN = (1, 0)$ conformal superspace in six dimensions} \label{geometry}

Here we collect the essential details of the 
superspace geometry of \cite{BKNT}. We refer the reader to appendix A of \cite{BKNT} for our notation 
and conventions.

We begin with a curved six-dimensional $\cN=(1,0)$ superspace
 $\cM^{6|8}$ parametrized by
local bosonic $(x^m)$ and fermionic coordinates $(\theta_i)$:
\be z^M = (x^m, \ \q^\mu_i) \ ,
\ee
where $m = 0, 1, \cdots, 5$, $\mu = 1, \cdots , 4$ and $i = 1, 2$. The 
structure group is chosen to be the full 6D $\cN = (1, 0)$ superconformal group and 
the covariant derivatives are postulated to have the form
\be
\nabla_A  
= E_A - \hf \Omega_A{}^{ab} M_{ab} - \Phi_A{}^{kl} J_{kl} - B_A \mathbb D
	- \mathfrak{F}_A{}_B K^B \ .
\ee
Here $E_A = E_A{}^M \partial_M$ is the inverse vielbein, 
$M_{ab}$ are the Lorentz generators, $J^{ij}$ are generators of the 
$\rm SU(2)$ group, $\mathbb D$ is the dilatation generator and $K^A = (K^a , S^\a_i)$ are the special superconformal 
generators. We associate the Lorentz $\Omega_A{}^{ab}$, SU(2) $\Phi_A{}^{kl}$, 
dilatation $B_A$ and special conformal $\frak{F}_{AB}$ 
connections with their respective generators.

The Lorentz generators obey
\bsubeq \label{SCA}
\begin{align}
[M_{ab} , M_{cd}] &= 2 \eta_{c[a} M_{b] d} - 2 \eta_{d [a} M_{b] c} \ , \\
[M_{ab} , \nabla_c ] &= 2 \eta_{c [a} \nabla_{b]} \ , \\
 [M_\a{}^\b , \nabla_\g^k] &= - \d_\g^\b \nabla_\a^k + \frac{1}{4} \d^\b_\a \nabla_\g^k
 \ .
\end{align}
The $\rm SU(2)$ and dilatation generators obey
\begin{align}
[J^{ij}, J^{kl}] &= \eps^{k(i} J^{j)l} + \eps^{l(i} J^{j)k} \ , \quad [J^{ij} , \nabla_\a^k] = \eps^{k(i} \nabla_\a^{j)} \ ,  \\
[\mathbb D , \nabla_a] &= \nabla_a \ , \quad [\mathbb D , \nabla_\a^i] = \hf \nabla_\a^i \ .
\end{align}
The Lorentz and $\rm SU(2)$ 
generators
act
on the special conformal generators $K^A$ as
\begin{align}
[M_{ab} , K_c] &= 2 \eta_{c[a} K_{b]} \ , \quad 
  [M_\a{}^\b , S^\g_k] = \d^\g_\a S^\b_k - \frac{1}{4} \d^\b_\a S^\g_k
\ , \\
[J^{ij} , S^\g_k] &= \d_k^{(i} S^{\g j)} \ ,
\end{align}
while the dilatation generator acts on  $K_A$ as
\begin{align}
[\mathbb D , K_a] = - K_a \ , \quad [\mathbb D, S^\a_i] &= - \hf S^\a_i \ .
\end{align}
Among themselves, the generators $K_A$ obey the algebra
\begin{align}
\{ S^\a_i , S^\b_j \} = - 2 \ri \eps_{ij} (\tilde{\g}^c)^{\a\b} K_c \ .
\end{align}
Finally, the algebra of $K_A$ with $\nabla_A$ is given by
\begin{align}
[K_a , \nabla_b] &= 2 \eta_{ab} \mathbb D + 2 M_{ab} \ , \quad
[K_a , \nabla_\a^i ] = - \ri (\g_a)_{\a\b} S^{\b i} \ , \\
[S^\a_i , \nabla_a] &= - \ri (\tilde{\g}_a)^{\a\b} \nabla_{\b i} \ , \quad
\{ S^\a_i , \nabla_\b^j \} = 2 \d^\a_\b \d^j_i \mathbb D - 4 \d^j_i M_\b{}^\a + 8 \d^\a_\b J_i{}^j \ .
\end{align}
\esubeq

The covariant derivatives obey (anti-)commutation relations of the form
\begin{align}
[ \nabla_A , \nabla_B \}
	&= -T_{AB}{}^C \nabla_C
	- \frac{1}{2} \RM_{AB}{}^{cd} M_{cd}
	- \RJ_{AB}{}^{kl} J_{kl}
	\non \\ & \quad
	- \RD_{AB} \mathbb D
	- \RS^{\phantom{k}}_{AB}{}_\g^k S^\g_k
	- \RK^{\phantom{k}}_{ABc} K^c~, \label{nablanabla}
\end{align}
where $T_{AB}{}^C$ is the torsion, and $\RM_{AB}{}^{cd}$, $\RJ_{AB}{}^{kl}$, $\RD_{AB}$, $\RS_{AB}{}_\g^K$ and $\RK_{AB}{}_c$ 
are the curvatures corresponding to the Lorentz, $\rm SU(2)$, dilatation, $S$-supersymmetry and special conformal boosts, 
respectively.

The full gauge group  of conformal supergravity, $\cG$, 
is generated by 
{\it covariant general coordinate transformations}, 
$\delta_{\rm cgct}$, associated with a parameter $\xi^A$ and 
{\it standard superconformal transformations}, 
$\delta_{\cH}$, associated with a parameter $\L^{\ul a}$. 
The latter include
the dilatation,
Lorentz, 
$\rm SU(2)$, 
and special conformal
(bosonic and fermionic) transformations.
The covariant derivatives transform as
\bea 
\d_\cG \nabla_A &=& [\cK , \nabla_A] \ ,
\label{TransCD}
\eea
where $\cK$ denotes the first-order differential operator
\bea
\cK = \xi^C \nabla_C + \hf \L^{ab} M_{ab} + \L^{ij} J_{ij} + \L \mathbb D + \L_A K^A ~.
\eea
Covariant (or tensor) superfields transform as
\be
\d_{\cG} T = \cK T~.
\ee

To describe conformal supergravity, the covariant derivative 
algebra \eqref{nablanabla} must be constrained as \cite{BKNT}
\bsubeq
\bea
\{ \nabla_\a^i , \nabla_\b^j \} &=& - 2 \ri \eps^{ij} (\g^a)_{\a\b} \nabla_a \ , \\ 
\left[ \nabla_a , \nabla_\a^i \right] &=& (\g_a)_{\a\b} \Big(
W^{\b\g} \nabla_\g^i
+ 4 \ri\, X_\d^i{}^{\b\g} M_\g{}^\d
- \frac{\ri}{2} X^{\g i} M_\g{}^\b
- 5\ri\, X^\b_j J^{ij}
+ \frac{5\ri}{4} X^{\b i} \mathbb D \non\\
&&
+ \frac{\ri}{4} Y_\g{}^\b{}^{ij} S^\g_j
+ \frac{\ri}{4} \nabla_{\g\d} W^{\d\b} S^{\g i}
- \frac{5\ri}{16} Y S^{\b i} \non\\
&&
+ \frac{\ri}{3} (\g^{bc})_\d{}^\g \big( \nabla^{\phantom{i}}_b X_\g^i{}^{\d \b} 
- \frac{3}{4} \d^\b_\g \nabla^{\phantom{k}}_b X^{\d i} \big) K_c
\Big) \ ,
\eea
\esubeq
where $W^{\a\b}$ is the super-Weyl tensor, which satisfies
\be W^{\a\b} = W^{\b\a} \ , \quad S^\g_k W^{\a\b} = 0 \ , \quad \mathbb D W^{\a\b} = W^{\a\b} \ ,
\ee
and the Bianchi identities
\bsubeq
\bea 
\nabla_\a^{(i} \nabla_{\b}^{j)} W^{\g\d} &=& - \d^{(\g}_{[\a} \nabla_{\b]}^{(i} \nabla_{\r}^{j)} W^{\d) \r} \ , \\
\nabla_\a^k \nabla_{\g k} W^{\b\g} - \frac{1}{4} \d^\b_\a \nabla_\g^k \nabla^{\phantom{k}}_{\d k} W^{\g\d}
 &=& 8 \ri \nabla_{\a \g} W^{\g \b} \ .
\label{WBI}
\eea
\esubeq
Here the descendents of $W^{\a\b}$ are defined as
\bsubeq
\begin{align}
X_\g^k{}^{\a\b} &:= -\frac{\ri}{4} \nabla_\g^k W^{\a\b} - \d^{(\a}_{\g} X^{\b) k} \ , \quad X^{\a i} := -\frac{\ri}{10} \nabla_{\b}^i W^{\a\b} \ , \\
Y_\a{}^\b{}^{ij} &:= - \frac{5}{2} \Big( \nabla_\a^{(i} X^{\b j)} - \frac{1}{4} \d^\b_\a \nabla_\g^{(i} X^{\g j)} \Big)
= - \frac{5}{2} \nabla_\a^{(i} X^{\b j)} \ , \\
Y &:= \frac{1}{4} \nabla_\g^k X^\g_k \ , \\
Y_{\a\b}{}^{\g\d} &:=
\nabla_{(\a}^k X^{\phantom{k}}_{\b) k}{}^{\g\d}
- \frac{1}{6} \d_\b^{(\g} \nabla_\r^k X^{\phantom{k}}_{\a k}{}^{\d) \r}
- \frac{1}{6} \d_\a^{(\g} \nabla_\r^k X^{\phantom{k}}_{\b k}{}^{\d) \r}
\ .
\end{align}
\esubeq
Note that $X_\g^k{}^{\a\b}$ is traceless, $Y_\a{}^{\b\, ij}$ is symmetric in its $\rm SU(2)$ indices and traceless in its
spinor indices, and $Y_{\a\b}{}^{\g\d}$ is separately symmetric in its
upper and lower spinor indices and traceless.

Upon taking a spinor covariant derivative of the descendent fields one finds
\bsubeq \label{eq:Wdervs}
\bea
\nabla_\a^i X^{\b j} &=& - \frac{2}{5} Y_\a{}^\b{}^{ij}
	- \frac{2}{5} \eps^{ij} \nabla_{\a \g} W^{\g\b}
- \hf \eps^{ij} \d_\a^\b Y \ , \\
\nabla_\a^i X_\b^j{}^{\g\d}
&=& 
\hf \d^{(\g}_\a Y_\b{}^{\d)}{}^{ij} - \frac{1}{10} \d_\b^{(\g} Y_\a{}^{\d)}{}^{ij}
- \hf \eps^{ij} Y_{\a\b}{}^{\g\d}
- \frac{1}{4} \eps^{ij} \nabla_{\a\b} W^{\g\d} \non\\
&&+ \frac{3}{20} \eps^{ij} \d_\b^{(\g} \nabla_{\a \r} W^{\d) \r}
- \frac{1}{4} \eps^{ij} \d^{(\g}_\a \nabla_{\b \r} W^{\d) \r} \ , \\
\nabla_\a^i Y &=& - 2 \ri \nabla_{\a \b} X^{\b i} \ , \\
\nabla_\g^k Y^{\phantom{k}}_{\a}{}^\b{}^{ij} &=& 
\frac{2}{3} \eps^{k (i} \Big( 
- 8 \ri \nabla_{\g\d} X_\a^{j)}{}^{\d \b}
- 4 \ri \nabla_{\a\d} X_\g^{j)}{}^{\d\b}
+ 3 \ri \nabla_{\g\a} X^{\b j)} \non\\
&&+ 3 \ri \d^\b_\g \nabla_{\a \d} X^{\d j)}
- \frac{3 \ri}{2} \d^\b_\a \nabla_{\g\d} X^{\d j)}
\Big) \ , \\
\nabla_\e^i Y^{\phantom{i}}_{\a\b}{}^{\g\d} &=& 
- 4 \ri \nabla_{\e (\a} X_{\b)}^l{}^{\g\d}
+ \frac{4 \ri}{3} \d^{(\g}_{(\a} \nabla_{\b) \r} X_\e^l{}^{\d ) \r} 
+ \frac{8 \ri}{3} \d^{(\g}_{(\a} \nabla_{|\e \r|} X_{\b)}^l{}^{\d) \r} \non\\
&&+ 8 \ri \d^{(\g}_{\e} \nabla_{\r (\a} X_{\b)}^l{}^{\d) \r} \ .
\eea
\esubeq
The descendant superfields transform under $S$-supersymmetry as follows:
\bsubeq
\begin{align}
S^\a_i X_\b^j{}{}^{\g\d} &= -\ri \,\d^j_i \d^\a_\b W^{\g\d} 
+ \frac{2 \ri}{5} \d^j_i \d^{(\g}_{\b} W^{\d) \a} , \qquad
S^\a_i X^{\b j} = \frac{8\ri}{5} \d_i^j W^{\a\b} ~, \\
S^\g_k Y^{\phantom{\gamma}}_\a{}^{\b}{}^{ij} &= \d^{(i}_k \Big( 
- 16 X_\a^{j)}{}^{\g\b} + 2 \d_\a^\b X^{\g j)}
- 8 \d^\g_\a X^{\b j)} \Big) \ , \\
S^\r_j Y_{\a\b}{}^{\g\d} &= 24 \Big( \d^\r_{(\a} X^{\phantom{\rho}}_{\b) j}{}^{\g\d}
- \frac{1}{3} \d^{(\g}_{(\a} X^{\phantom{\gamma}}_{\b) j}{}^{\d) \r} \Big) \ , \quad S^\a_i Y = - 4 X^\a_i \ .
\end{align}
\esubeq

%%%%%%%%%%%%%%%%%%%%%%%%%%%%%%%%%%%%%%%%%%%%%%%%%%%%%%
%%%%%%%%%%%%%%%%%%%%%%%%%%%%%%%%%%%%%%%%%%%%%%%%%%%%%%

\section{The prepotential for the $\cO(2)$ multiplet} \label{PrepotO(2)}

The prepotential formulation for the $\cO(2)$ multiplet was first given 
in Minkowski superspace in \cite{GL85}, where it was shown that 
the prepotential is a spinor superfield $\r_\a^i$.  In this apppendix we extend the construction 
in \cite{GL85} to supergravity by making use of conformal superspace. 

We will work in projective superspace, where the supermanifold $\cM^{6|8}$ is augmented
with an additional $\rm \mathbb CP^1$ parametrized by an isotwistor
coordinate $v^i \in \mathbb C^2 \setminus \{0\}$.
Matter fields are constructed in terms of covariant projective multiplets
$\cQ^{(n)}(z,v)$, which are holomorphic in the isotwistor $v^i$
and of definite homogeneity, $\cQ^{(n)}(z, cv) = c^n \cQ^{(n)}(z, v)$,
on an open domain of $\mathbb C^2 \setminus \{0\}$. Such superfields
are intrinsically defined on $\rm \mathbb CP^1$.

It is useful to introduce an additional fixed isotwistor $u_i$ which obeys
$v^i u_i \neq 0$. Given a superfield $T^{i_i \cdots i_n}$ with symmetric SU(2) indices $T^{i_i \cdots i_n} = T^{(i_i \cdots i_n)}$ 
(and suppressed Lorentz indices) we define
\be
T^{(m)(m - n)} := v_{i_1} \cdots v_{i_m} \frac{u_{i_{m+1}}}{(v , u)} \cdots \frac{u_{i_{n}}}{(v , u)} T^{i_1 \cdots i_n} \ , 
\quad (v, u) := v^k u_k \ .
\ee
We also introduce
\bea
\nabla_\a^{(1)} := v_i \nabla_\a^i \ , \quad \nabla_\a^{(-1)} = \frac{u_i}{(v,u)} \nabla_\a^i
\eea
and the following derivative operations
\begin{align}
\partial^{(2)} = (v,  u) v_i\frac{\partial}{\partial u_i}~,\qquad
\partial^{(0)} = v^i \frac{\partial}{\partial v^i} - u_i \frac{\pa}{\pa u_i}~,\qquad
\partial^{(-2)} = \frac{1}{(v, u)} u^i\frac{\partial}{\partial v^i}~.
\end{align}
Fields and operators
of definite homogeneity in $v^i$ can be interpreted as possessing
definite $\partial^{(0)}$ charge. Note that one can express the SU(2) generator in terms of the above derivative 
operators as follows
\begin{align}
J_{ij} = - v_i v_j \partial^{(-2)} + \frac{v_{(i} u_{j)}}{(v, u)} \partial^{(0)} + \frac{1}{(v, u)^2}  u_i u_j \partial^{(2)}~.
\end{align}

Let us now use the above isotwistor notation to write down a candidate for the prepotential description 
of the $\cO(2)$ multiplet, $G^{(2)} = v_i v_j G^{ij}$, which satisfies the constraint
\be \nabla_\a^{(1)} G^{(2)} = 0 \ .
\ee

Considering the generalisation of the result in \cite{GL85} to supergravity, one is led to the following 
natural ansatz
\be G^{(2)} = \nabla^{(4)} \Big( \nabla^{\a (-3)} \r_\a^{(1)} + a \ri \nabla_\a^{(-1)} \nabla^{\a\b} \r_\b^{(-1)} 
+ b \ri \nabla_\a^{(-1)} (W^{\a\b} \r_\b^{(-1)} ) \Big) \ ,
\ee
where $a$ and $b$ are some coefficients to be determined and $\r_\a^i$ is the prepotential which we assume to be primary and 
of dimension 1/2. 
Here we have introduced the covariant differential operators
\bea 
\nabla^{(4)} := \frac{1}{4!} \eps^{\a\b\g\d} \nabla_\a^{(1)} \nabla_\b^{(1)} \nabla_\g^{(1)} \nabla_\d^{(1)} \ , \quad
\nabla^{\a (3)} := \frac{1}{3!} \eps^{\a\b\g\d} \nabla_\a^{(1)} \nabla_\b^{(1)} \nabla_\g^{(1)} \ .
\eea

For consistency we require independence of the isotwister $u^i$, which amounts to requiring
\be \partial^{(2)} G^{(2)} = 0 \ .
\ee
Using the property $\nabla_\a^{(1)} \nabla^{(4)} = \nabla^{(4)} \nabla_\a^{(1)} = 0$ and the above requirement fixes the 
coefficients $a$ and $b$ as
\be G^{(2)} = \nabla^{(4)} \Big( \nabla^{\a (-3)} \r_\a^{(1)} - 2 \ri \nabla_\a^{(-1)} \nabla^{\a\b} \r_\b^{(-1)}
- 8 \ri \nabla_\a^{(-1)} (W^{\a\b} \r_\b^{(-1)} ) \Big) \ .
\ee

The prepotential is defined up to some gauge freedom. Specifically, $G^{(2)}$ is 
unchanged by the following shift in the prepotential
\be \r_\a^i \rightarrow \r'{}_\a^i = \r_\a^i + \nabla_\a^i \t + \nabla_{\b j} \t_\a{}^\b{}^{ij} \ , \quad \t_\a{}^\a{}^{ij} = 0 \ , \quad 
\t_\a{}^{\b}{}^{ij} = \t_\a{}^{\b}{}^{(ij)} \ , \label{RhoPrepotGaugeTrans}
\ee
where $\t$ and $\t_\a{}^\b{}^{ij}$ are dimensionless primary superfields. 
It is rather simple to show that the transformations associated with the 
scalar $\t$ leave $G^{(2)}$ invariant since
\be \{ \nabla_\a^{(1)} , \nabla^{\a (-3)} \} \t = 2 \ri \nabla_\a^{(-1)} \nabla^{\a\b} \nabla_\b^{(-1)} \t + 8 \ri \nabla_\a^{(-1)} (W^{\a\b} \nabla_\b^{(-)} \t) \ .
\ee
To show that $G^{(2)}$ is also invariant with respect to the transformations associated with $\t_\a{}^\b{}^{ij}$ one 
uses the identities
\bsubeq
\begin{align}
\nabla^{(4)} \nabla_\a^{(1)} &= 0 \ , \\
\nabla^{\a (-3)} \nabla_\b^{(-1)} &= \d^\a_\b \nabla^{(-4)} \ , \\
\{ \nabla_\a^{(-1)} \nabla^{\a\b} , \nabla_\g^{(1)} \} \t_\b{}^\g{}^{(-2)}
&= - 4 \nabla_\a^{(-1)} (W^{\a\b} \nabla_\g^{(1)} \t_\b{}^{\g (-2)} ) \ , \\
\{ \nabla^{\a (-3)} , \nabla_\b^{(1)} \} \t_\a{}^\b{}^{(1)(-1)} &= - 2 \ri \nabla_\a^{(-1)} \nabla^{\a\b} \nabla_\g^{(-1)} \t_\b{}^\g{}^{(1)(-1)} \non\\
&\quad \, -8\ri \nabla_\a^{(-1)} (W^{\a\b} \nabla_\g^{(-1)} \t_\b{}^\g{}^{(1)(-1)}) \ .
\end{align}
\esubeq

We now have to check that $G^{(2)}$ is primary. To do so, we apply the following identities
\bsubeq
\bea
\left[ S^{\a (-1)} , \nabla^{(4)} \right] 
&=& - 12 \nabla^{\a (3)} + \nabla^{\b (3)} (2 \d^\a_\b \mathbb D - 4 M_\b{}^\a - 4 \d^\a_\b \partial^{(0)}) \ , \\
\{ S^{\a (-1)} , \nabla^{\b (-3)} \}
&=& 4 \eps^{\a\b\g\d} \nabla_\g^{(-1)} \nabla_\d^{(-1)} \partial^{(-2)} \ , \\
\{ S^{\a (-1)} , \nabla_\g^{(-1)} \nabla^{\g \b} \}
&=& - 8 \nabla^{\a\b} \partial^{(-2)} - 2 \ri \eps^{\a\b\g\d} \nabla_\g^{(-1)} \nabla_\d^{(-1)} \ 
\eea
\esubeq
to show
\be
S^{\a (-1)} G^{(2)} = 0 \ .
\ee
Furthermore, since $G^{(2)}$ is independent of the isotwistor $u^i$ we must have
\be
S^\a_i G^{jk} = 0 \ ,
\ee
and thus $G^{ij}$ is primary.

Finally, since $G^{(2)}$ is independent of the isotwistor $u^i$, $G^{ij}$ can be written without isotwistors 
as follows
\be G^{ij} = \hf \nabla^{(ijkl} \nabla_{\a klp} \r_\a^{p)} - \frac{6 \ri}{5} \nabla^{ijkl} \big( \nabla_{\a k} \nabla^{\a\b} \r_{\b l} 
+ 4 \nabla_{\a k} (W^{\a\b} \r_{\b l}) \big) \ .
\ee 

%%%%%%%%%%%%%%%%%%%%%%%%%%%%%%%%%%%%%%%%%%%%%%%%%%%%%%
%%%%%%%%%%%%%%%%%%%%%%%%%%%%%%%%%%%%%%%%%%%%%%%%%%%%%%

\section{The Yang-Mills multiplet in conformal superspace} \label{YMmultiplet}

In this appendix, we give the results needed for the description of the Yang-Mills multiplet in conformal 
superspace. It is based almost verbatim on appendix C of \cite{BKNT}.

To describe a non-abelian vector multiplet, the covariant derivative $\nabla_A $ 
is replaced with a gauge covariant one, 
\bea
{\bm\nabla}_A := \nabla_A 
- \ri V_A
~.
\label{SYM-derivatives}
\eea
Here the  gauge connection one-form $V_A$ 
takes its values in the Lie algebra 
of the (unitary) Yang-Mills gauge group, $G_{\rm YM}$, with its (Hermitian) generators 
commuting with all the generators of the superconformal algebra. 
The algebra of the gauge covariant derivatives is
\bea
[{\bm \nabla}_A, {\bm \nabla}_B \} 
&=&
 -T_{AB}{}^C{\bm \nabla}_C
-\hf \RM_{AB}{}^{cd} M_{cd}
-\RJ_{AB}{}^{kl} J_{kl}
- \RD_{AB} \mathbb D 
\non\\
&&
 - \RS^{\phantom{\gamma}}_{AB}{}^\g_k S_\g^k
	- \RK_{AB}{}^c K_c
	- \ri F_{AB} \ ,
\eea
where the torsion and curvatures are those of conformal superspace 
and $F_{AB}$ is  the gauge covariant field strength two-form. 
It satisfies the Bianchi identity
\be 
\bm \nabla_{[A} F_{BC\}} 
+ T_{[AB}{}^D F_{|D| C\}} = 0
~. \label{FBI45}
\ee
The Yang-Mills gauge transformation acts on the gauge covariant 
derivatives $\bm \nabla_A$ and a matter  superfield $U$ (transforming 
in some representation of the gauge group) 
as
\be 
\bm \nabla_A ~\rightarrow~ \re^{\ri  \t} \bm \nabla_A \re^{- \ri \t } , 
\qquad  U~\rightarrow~ U' = \re^{\ri  \t} U~, 
\qquad \t^\dag = \t \ ,
\label{2.2}
\ee
where the Hermitian gauge parameter ${\t} (z)$ takes its values in the Lie algebra 
of $G_{\rm YM}$. 

Some components of the superform field strength have to be constrained 
in order to describe an irreducible multiplet. In conformal superspace, the components of 
the field strength are constrained as
\bsubeq
\bea 
F_\a^i{}_\b^j &=& 0 \ , \\ 
F^{\phantom{j}}_a{}_\b^j &=& (\g_a)_{\a\b} W^{\b i} \ , \label{YMsuperformFa} \\
F_{ab} &=& - \frac{\ri}{8} (\g_{ab})_\b{}^\a \bm \nabla_\a^k W^\b_k \ , 
\eea
\esubeq
where $W^{\a i}$ is a conformal primary of dimension $\frac{3}{2}$,
 $S^\g_k W^{\a i} =0$ and
 $\bbD W^{\a i}= \frac{3}{2} W^{\a i}$. The Bianchi identity \eqref{FBI45} 
constrains $W^{\a i}$ to obey the differential constraints
\bea 
\bm \nabla_\g^k W^\g_k = 0 \ , \quad \bm \nabla_\a^{(i} W^{\b j)}
= \frac{1}{4} \d_\a^\b \bm \nabla_\g^{(i} W^{\g j)}
\label{vector-Bianchies} \ .
\eea

It is useful to introduce the following descendant superfield:
\begin{align}
X^{ij} := \frac{\ri}{4} \bm\nabla^{(i}_\g W^{\g j)} \ .
\label{C.8}
\end{align}
The superfield $W^{\a i}$ and $X^{ij}$, together with
\be 
F_\a{}^\b = - \frac{\ri}{4} \Big( \bm \nabla_\a^k W^\b_k - \frac{1}{4} \d_\a^\b \bm \nabla_\g^k W^\g_k \Big)
= - \frac{\ri}{4} \bm \nabla_\a^k W^\b_k \ ,
\ee
satisfy the following identities:
\bsubeq \label{VMIdentities}
\bea
\bm\nabla_\a^i W^{\b j}
&=& 
- \ri \d_\a^\b X^{ij} - 2 \ri \eps^{ij} F_\a{}^\b \ , \\
\bm \nabla_\a^i F_\b{}^\g
&=&
- \bm \nabla_{\a\b} W^{\g i}
- \d^\g_\a \bm \nabla_{\b\d} W^{\d i}
+ \frac{1}{2} \d_\b^\g \bm \nabla_{\a \d} W^{\d i}
\ , \\
\bm\nabla_\a^i X^{jk}
&=&
2 \eps^{i(j} \bm \nabla_{\a\b} W^{\b k)} \ , \\
S^\g_k F_\a{}^\b &=& - 4 \ri \d^\g_\a W^\b_k
+ \ri \d^\b_\a W^\g_k \ , \qquad
S^\g_k X^{ij} = - 4 \ri \d_k^{(i} W_{\phantom{k}}^{\g j)} \ .
\eea
\esubeq

%%%%%%%%%%%%%%%%%%%%%%%%%%%%%%%%%%%%%%%%%%%%%%%%%%%%%%
%%%%%%%%%%%%%%%%%%%%%%%%%%%%%%%%%%%%%%%%%%%%%%%%%%%%%%

\section{A superform description for the $\cN = (2, 0)$ tensor multiplet and its deformation} \label{c4Form}

We give the superform description for the $\cN = (2, 0)$ tensor multiplet and introduce 
a closed four-form by deforming the constraints defining the tensor multiplet.

The tensor multiplet can be described by a two-form gauge potential. The field strength three-form 
$H_3 = \frac{1}{3!} \rd z^C \rd z^B \rd z^A H_{ABC}$ is given in terms of its two-form gauge potential $B_2 = \frac{1}{2} \rd z^B \rd z^A B_{AB}$ by
\be H_3 = \rd B_2 \implies H_{ABC} = 3 D_{[A} B_{BC\}} + 3 T_{[AB}{}^D B_{|D|C\}} \ ,
\ee
where the only non-vanishing component of the torsion is
\be
T_\a^{\hi }{}_\b^{\hj}{}^a = 2 \ri \Omega^{\hi \hj} (\g^a)_{\a\b} \ .
\ee
The existence of the gauge potential requires that the Bianchi identity
\be \rd H_3 = 0 \implies D_{[A} H_{BCD\}} + \frac{3}{2} T_{[AB}{}^{E} H_{|E|CD\}} = 0
\ee
be satisfied. To describe the tensor multiplet, one must impose the 
following constraints on the lowest components of the 
superform field strength:
\bsubeq
\be
H_\a^{\hi }{}_\b^{\hj}{}_\g^{\hk} = 0 \ , \quad 
H_a^{\phantom{i}}{}_\a^{\hi }{}_\b^{\hj} = 2 \ri (\g_a)_{\a\b} \Phi^{\hi \hj} 
\ , \quad \Phi^{(\hi \hj)} = \Phi^{\hi \hj} \Omega_{\hi \hj} = 0 \ .
\ee
The Bianchi identities for $H_3$ can then be solved giving the remaining components:
\bea
H^{\phantom{i}}_{ab}{}_\a^{\hi } &=& - \frac{1}{4} (\g_{ab})_\a{}^\b \l_\b^{\hi } \ , \\
H_{abc} &=& \frac{1}{8} (\tilde{\g}_{abc})^{\a\b} H_{\a\b} \ , \quad H_{[\a\b]} = 0 \ ,
\eea
\esubeq
where $\Phi^{\hi \hj}$ is required to satisfy the differential constraint
\be D_\a^{\hi } \Phi^{\hj\hk} - \Omega^{\hi [\hj} \l_\a^{\hk]} 
- \frac{1}{4} \Omega^{\hj\hk} \l_\a^{\hi } = 0 \ , \label{defDiffConstTensor}
\ee
and its corollaries
\bsubeq
\bea 
D_\a^{\hi } \l_\b^{\hj} &=& 2 \ri \Omega^{\hi \hj} H_{\a\b} + 4 \ri \partial_{\a\b} \Phi^{\hi \hj} \ , \\
D_\a^{\hi} H^{\phantom{i}}_{\b\g} &=& \partial^{\phantom{i}}_{\a (\b} \l_{\g)}^{\hi} \ , \quad \partial^{\a\b} \l_\b^{\hi} = 0 \ .
\eea
\esubeq
The constraint \eqref{defDiffConstTensor} is the defining constraint for the $\cN = (2, 0)$ tensor multiplet.

We now wish to describe a closed four-form $H_4 = \frac{1}{4!} \rd z^D \rd z^C \rd z^B \rd z^A H_{ABCD}$, which satisfies 
the closure condition
\be \rd H_4 = 0 \implies D_{[A} H_{BCDE\}} + 2 T_{[AB}{}^{F} H_{|F|CDE\}} = 0 \ . \label{closure4form}
\ee
To do this we proceed by obstructing the constraint defining the tensor multiplet by a 
closed 4-form $H_4$ as\footnote{A similar procedure of obstructing the closure condition of a $p$-superform to 
obtain a closed $(p+1)$-superform was used in \cite{ALR14} to construct a chain of 
closed $\cN = (1, 0)$ superforms.}
\be \rd H_3 = H_4 \quad \implies \quad 4 D_{[A} H_{BCD\}} + 6 T_{[AB}{}^{E} H_{|E|CD\}} = H_{ABCD} \label{H3H4superform}
\ee
such that the constraint on the tensor multiplet is deformed to
\be D_\a^{\hi } \Phi^{\hj\hk} - \Omega^{\hi [\hj} \l_\a^{\hk]} 
- \frac{1}{4} \Omega^{\hj\hk} \l_\a^{\hi } = H_\a^{\hi }{}^{, \hj\hk} \ , \label{tensorDeform}
\ee
where $H_\a^{\hi}{}^{, \hj\hk} = H_\a^{\hi}{}^{, [\hj\hk]} $ is completely traceless and the 
four-form is constructed completely in terms of $H_\a^{\hi}{}^{, \hj\hk}$.
The first non-vanishing component of $H_4$ is fixed 
by the condition \eqref{H3H4superform} and the constraint \eqref{tensorDeform} to be
\bea
H^{\phantom{i}}_a{}_\a^{\hi }{}_\b^{\hj}{}_\g^{\hk} &=& -2 \ri (\g_a)_{\a\b} H_\g^{\hk}{}^{,\hi \hj} 
- 2 \ri (\g_a)_{\b\g} H_\a^{\hi }{}^{,\hj\hk} 
- 2 \ri (\g_a)_{\g\a} H_\b^{\hj}{}^{,\hk\hi }  \ .
\eea
The remaining components can be determined by finding the conditions 
that follow from the constraint \eqref{tensorDeform} and using eq. \eqref{H3H4superform}. The
consistency conditions on $H_\a^{\hi}{}^{, \hj\hk}$ that follow from the constraint \eqref{tensorDeform} 
will give the requirements for $H_4$ to be closed.

By taking successive spinor derivatives of 
the superfield $\Phi^{\hi\hj}$ one finds the following results at dimension 4:
\bsubeq
\begin{align} D_\a^{\hi} \l_\b^{\hj} &= H_{\a\b}{}^{\hi\hj}
+ 2 \ri \Omega^{\hi\hj} H_{\a\b} + 4 \ri \partial_{\a\b} \Phi^{\hi\hj} \ , \quad \Omega_{\hi\hj} H_{\a\b}{}^{\hi\hj} = 0 \ , \\
D_\a^{\hi} H_\b^{\hj}{}^{,\hk\hl} + D_\b^{\hj} H_\a^{\hi}{}^{,\hk\hl}
&=
- \Omega^{\hi[\hk} H_{\b\a}{}^{|\hj|\hl]}
- \Omega^{\hj[\hk} H_{\a\b}{}^{|\hi|\hl]}
- \frac{1}{4} \Omega^{\hk\hl} H_{\a\b}{}^{\hi\hj}
- \frac{1}{4} \Omega^{\hk\hl} H_{\b\a}{}^{\hj\hi}
\ ,
\end{align}
\esubeq
while at dimension 9/2 one finds
\bsubeq
\begin{align}
D_\g^{\hi} H_{\a\b}
&= \partial_{\g (\a} \l_{\b)}^{\hi} 
- \frac{\ri}{30} D_{\g \hk} H_{(\a\b)}{}^{\hi\hk}
+ \frac{2 \ri}{15} D_{(\a \hk} H_{|\g|\b)}{}^{\hi\hk}\ , \\
15 \ri \partial^{\phantom{i}}_{[\a\b} \l_{\g]}^i
&= 
D_{\g \hk} H_{[\a\b]}{}^{\hi\hk}
- 4 D_{[\a \hk} H_{|\g|\b]}{}^{\hi\hk}
\ , \\
30 \ri \partial^{\phantom{i}}_{[\a\b} \l_{\g]}^{\hi}
&= D_{[\g \hk} H_{\a]\b}{}^{\hi\hk} - 3 D_{[\g \hk} H_{\a]\b}{}^{\hk\hi} \ , \\
0 &= 
D_\a^{\hi} H^{\phantom{i}}_{\b\g}{}^{\hj\hk}
+ 4 \ri \partial^{\phantom{i}}_{\b\g} H_\a^{\hi}{}^{,\hj\hk}
+ \Omega^{\hi\hk} \xi^{\phantom{i}}_{\a\b\g}{}^{\hj}
+ \Omega^{\hi\hj} \eps^{\phantom{i}}_{\a\b\g\d} \tilde{\xi}^{\d \hk}
+ (\underline{\a} \leftrightarrow \underline{\b})
\ .
\end{align}
Finally, at dimension 5 one finds
\be 4 \partial_{[a} H_{bcd]} = 
\frac{1}{1920} (\tilde{\g}_{[a})^{\g\d} (\tilde{\g}_{bcd]})^{\a\b} D_{\g \hk} D_{\d \hl} H_{\a\b}{}^{\hk\hl}
- \frac{1}{480} (\tilde{\g}_{[a})^{\g\d} (\tilde{\g}_{bcd]})^{\a\b} D_{\g \hk} D_{\a \hl} H_{\d\b}{}^{\hk\hl} \ .
\ee
\esubeq
All constraints on $H_\a^{\hi}{}^{, \hj \hk}$ are encoded 
in the closure of the four-form $H_4$.

From the above results one can determine the components of the four-form:
\bsubeq
\begin{align}
H^{\phantom{i}}_{ab}{}_\a^{\hi }{}_\b^{\hj} &=
- \hf (\g_{ab})_{\a}{}^{\g} H_{\b \g}{}^{\hj \hi}
- \hf (\g_{ab})_{\b}{}^\g H_{\a\g}{}^{\hi \hj}
\ ,
\\
H^{\phantom{i}}_{abc}{}_\g^{\hk } &=
\frac{\ri}{240} (\tilde{\g}_{abc})^{\a\b}
\big( D_{\g \hj} H_{\a\b}{}^{\hk\hj} - 4 D_{\a \hj} H_{\g \b}{}^{\hk\hj}
\big) 
\ , \\
H_{abcd} &=
\frac{1}{1920} (\tilde{\g}_{[a})^{\g\d} (\tilde{\g}_{bcd]})^{\a\b} D_{\g \hk} D_{\d \hl} H_{\a\b}{}^{\hk\hl}
- \frac{1}{480} (\tilde{\g}_{[a})^{\g\d} (\tilde{\g}_{bcd]})^{\a\b} D_{\g \hk} D_{\a \hl} H_{\d\b}{}^{\hk\hl} \ .
\end{align}
\esubeq
It is important to emphasise that all the differential constraints on $H_{\a\b}{}^{\hi\hj}$ can be projected out of the 
closure condition \eqref{closure4form}. However, we do not give them explicitly here for simplicity.

%%%%%%%%%%%%%%%%%%%%%%%%%%%%%%%%%%%%%%%%%%%%%%%%%%%%%%
%%%%%%%%%%%%%%%%%%%%%%%%%%%%%%%%%%%%%%%%%%%%%%%%%%%%%%

\begin{footnotesize}

\end{footnotesize}


\begin{thebibliography}{66}


\bibitem{FZ}
 S.~Ferrara and B.~Zumino,
``Transformation properties of the supercurrent,''
Nucl.\ Phys.\  B {\bf 87}, 207 (1975).

\bibitem{KKT} 
  Y.~Korovin, S.~M.~Kuzenko and S.~Theisen,
  ``The conformal supercurrents in diverse dimensions and conserved superconformal currents,''
  JHEP {\bf 1605}, 134 (2016)
  [arXiv:1604.00488 [hep-th]].

\bibitem{HL}
  P.~S.~Howe and U.~Lindstr\"om,
  ``The supercurrent in five dimensions,''
  Phys.\ Lett.\ B {\bf 103}, 422 (1981).

\bibitem{HST83} 
  P.~S.~Howe, G.~Sierra and P.~K.~Townsend,
  ``Supersymmetry in six dimensions,''
  Nucl.\ Phys.\ B {\bf 221}, 331 (1983).

\bibitem{Gates} 
  S.~J.~Gates Jr.,
  ``Super $p$-form gauge superfields,''
  Nucl.\ Phys.\ B {\bf 184}, 381 (1981).
 
\bibitem{GGRS}
S.~J.~Gates Jr., M.~T.~Grisaru, M.~Ro\v{c}ek and W.~Siegel,
{\it Superspace, or One Thousand 
and One Lessons in Supersymmetry},
Benjamin/Cummings (Reading, MA),  1983, hep-th/0108200.

\bibitem{CPS} 
  T.~E.~Clark, O.~Piguet and K.~Sibold,
  ``Supercurrents, renormalization and anomalies,''
  Nucl.\ Phys.\ B {\bf 143}, 445 (1978).

\bibitem{new}
M.~F.~Sohnius and P.~C.~West,
``An alternative minimal off-shell version of N=1 supergravity,''
Phys.\ Lett.\  B {\bf 105}, 353 (1981).

\bibitem{GGS} 
  S.~J.~Gates Jr., M.~T.~Grisaru and W.~Siegel,
  ``Auxiliary field anomalies,''
  Nucl.\ Phys.\ B {\bf 203}, 189 (1982).

\bibitem{OS}
V.~Ogievetsky and E.~Sokatchev,
``On vector superfield generated by supercurrent,''
Nucl.\ Phys.\  B {\bf 124}, 309 (1977).

\bibitem{FZ2}
S.~Ferrara and B.~Zumino,
``Structure of linearized supergravity and conformal 
supergravity,''  Nucl.\ Phys.\  B {\bf 134}, 301 (1978).
  
\bibitem{Siegel77}
W.~Siegel, ``A derivation of the supercurrent superfield,''
Harvard  preprint  HUTP-77/A089 (December, 1977). 

\bibitem{old}
J.~Wess and B.~Zumino,
``Superfield Lagrangian for supergravity,''
Phys.\ Lett.\  B {\bf 74}, 51 (1978);
K.~S.~Stelle and P.~C.~West,
``Minimal auxiliary fields for supergravity,''
Phys.\ Lett.\  B {\bf 74},  330 (1978);
S.~Ferrara and P.~van Nieuwenhuizen,
``The auxiliary fields of supergravity,''
Phys.\ Lett.\  B {\bf 74}, 333 (1978).

\bibitem{Deser} 
S.~Deser, 
``Scale invariance and gravitational coupling,''
Annals Phys.\  {\bf 59}, 248 (1970).

\bibitem{Zumino} B. Zumino, 
``Effective Lagrangians and broken symmetries," 
in {\it Lectures on Elementary Particles and Quantum Field Theory,
Vol. 2}, S. Deser, M. Grisaru and H. Pendleton (Eds.), The M.I.T. Press, 
Cambridge, Mass. 1970, pp. 437-500.

\bibitem{Dirac} 
  P.~A.~M.~Dirac,
``Long range forces and broken symmetries,''
  Proc.\ Roy.\ Soc.\ Lond.\ A {\bf 333}, 403 (1973).

\bibitem{Siegel77_2}
W.~Siegel, ``A polynomial action for a massive, self-interacting chiral
  superfield coupled to supergravity,''
HUTP-77/A077 (1977)  .

\bibitem{KakuT} 
  M.~Kaku and P.~K.~Townsend,
  ``Poincar\'e supergravity as broken superconformal gravity,''
  Phys.\ Lett.\ B {\bf 76}, 54 (1978).
  
  \bibitem{MR03} 
  R.~Manvelyan and W.~R\"uhl,
  ``On the supermultiplet of anomalous currents in $d = 6$,''
  Phys.\ Lett.\ B {\bf 567}, 53 (2003)  [hep-th/0305138].
  
\bibitem{Stelle}
K.~S.~Stelle,
``Extended supercurrents and the ultraviolet finiteness of N=4 supersymmetric
Yang-Mills theory,'' in {\it Quantum Structure of Space and Time}, M. J. Duff and C. J. Isham (Eds.), 
Cambridge University Press, Cambridge, 1982, pp. 337--361.
  
\bibitem{Sohnius}
  M.~F.~Sohnius,
  ``The multiplet of currents for N=2 extended supersymmetry,''
  Phys.\ Lett.\  B {\bf 81}, 8 (1979).

\bibitem{HST}
P.~S.~Howe, K.~S.~Stelle and P.~K.~Townsend,
``Supercurrents,''  Nucl.\ Phys.\  B {\bf 192}, 332 (1981).
  
\bibitem{KT}
S.~M.~Kuzenko and S.~Theisen,
``Correlation functions of conserved currents in N = 2 superconformal
theory,''  Class.\ Quant.\ Grav.\  {\bf 17}, 665 (2000)  [hep-th/9907107].

\bibitem{BK10}
D.~Butter and S.~M.~Kuzenko,
``N=2 supergravity and supercurrents,''  JHEP {\bf 1012}, 080 (2010)
[arXiv:1011.0339 [hep-th]].

\bibitem{GIKOS}
A.~Galperin, E.~Ivanov, S.~Kalitzin, V.~Ogievetsky and E.~Sokatchev, 
``Unconstrained N=2 matter, Yang-Mills and supergravity theories in harmonic
superspace,'' Class.\ Quant.\ Grav.\  {\bf 1}, 469 (1984).

\bibitem{GIOS87}
A.~S.~Galperin, E.~A.~Ivanov, V.~I.~Ogievetsky and E.~Sokatchev,
``N=2 supergravity in superspace: Different versions and matter couplings,''
Class.\ Quant.\ Grav.\  {\bf 4}, 1255 (1987).

\bibitem{GIOS}
A.~S.~Galperin, E.~A.~Ivanov, V.~I.~Ogievetsky and E.~S.~Sokatchev,
{\it Harmonic Superspace}, Cambridge University Press, 
Cambridge, 2001.

\bibitem{BKNT} 
  D.~Butter, S.~M.~Kuzenko, J.~Novak and S.~Theisen,
  ``Invariants for minimal conformal supergravity in six dimensions,''
 JHEP {\bf 1612}, 072 (2016)
  [arXiv:1606.02921 [hep-th]].
  
\bibitem{KNS1} 
  S.~M.~Kuzenko, J.~Novak and I.~B.~Samsonov,
  ``The anomalous current multiplet in 6D minimal supersymmetry,''
  JHEP {\bf 1602}, 132 (2016)
  [arXiv:1511.06582 [hep-th]].
  
\bibitem{KNS2} 
  S.~M.~Kuzenko, J.~Novak and I.~B.~Samsonov,
  ``Chiral anomalies in six dimensions from harmonic superspace,''
  arXiv:1708.08238 [hep-th].

\bibitem{ISZ05} 
  E.~A.~Ivanov, A.~V.~Smilga and B.~M.~Zupnik,
  ``Renormalizable supersymmetric gauge theory in six dimensions,''
  Nucl.\ Phys.\ B {\bf 726}, 131 (2005)
  [hep-th/0505082].
  
\bibitem{GL85} 
  J.~Grundberg and U.~Lindstr\"om,
  ``Actions for linear multiplets in six dimensions,''
  Class.\ Quant.\ Grav.\  {\bf 2}, L33 (1985).
  
\bibitem{BSVanP} 
  E.~Bergshoeff, E.~Sezgin and A.~Van Proeyen,
  ``Superconformal tensor calculus and matter couplings in six dimensions,''
  Nucl.\ Phys.\ B {\bf 264}, 653 (1986)
  Erratum: [Nucl.\ Phys.\ B {\bf 598}, 667 (2001)].
  
\bibitem{LT-M12} 
  W.~D.~Linch, III and G.~Tartaglino-Mazzucchelli,
  ``Six-dimensional supergravity and projective superfields,''
  JHEP {\bf 1208}, 075 (2012)
  [arXiv:1204.4195 [hep-th]].
  
\bibitem{HST83-2} 
  P.~S.~Howe, K.~S.~Stelle and P.~K.~Townsend,
  ``The relaxed hypermultiplet: An unconstrained N=2 superfield theory,''
  Nucl.\ Phys.\ B {\bf 214}, 519 (1983).
  
\bibitem{ALR14} 
  C.~Arias, W.~D.~Linch, III and A.~K.~Ridgway,
  ``Superforms in six-dimensional superspace,''
  JHEP {\bf 1605}, 016 (2016)
  [arXiv:1402.4823 [hep-th]].

\bibitem{KNT17} 
  S.~M.~Kuzenko, J.~Novak and S.~Theisen,
  ``New superconformal multiplets and higher derivative invariants in six dimensions,''
  arXiv:1707.04445 [hep-th].
  
\bibitem{Sokatchev88} 
  E.~Sokatchev,
  ``Off-shell six-dimensional supergravity in harmonic superspace,''
  Class.\ Quant.\ Grav.\  {\bf 5}, 1459 (1988).
  
\bibitem{HU87} 
  P.~S.~Howe and A.~Umerski,
  ``Anomaly multiplets In six-dimensions and ten-dimensions,''
  Phys.\ Lett.\ B {\bf 198}, 57 (1987).
  
\bibitem{CvP11} 
  F.~Coomans and A.~Van Proeyen,
  ``Off-shell N=(1,0), D=6 supergravity from superconformal methods,''
  JHEP {\bf 1102}, 049 (2011)
  Erratum: [JHEP {\bf 1201}, 119 (2012)]
  [arXiv:1101.2403 [hep-th]].
  
\bibitem{BNT-M} 
  D.~Butter, J.~Novak and G.~Tartaglino-Mazzucchelli, 
``The component structure of conformal supergravity invariants in six dimensions,''
  JHEP {\bf 1705}, 133 (2017)
  [arXiv:1701.08163 [hep-th]].

\bibitem{BCSvP12} 
  E.~Bergshoeff, F.~Coomans, E.~Sezgin and A.~Van Proeyen,
  ``Higher derivative extension of 6D chiral gauged supergravity,''
  JHEP {\bf 1207}, 011 (2012)
  [arXiv:1203.2975 [hep-th]].
  
\bibitem{BPT}
  L.~Bonora, P.~Pasti and M.~Tonin,
``Cohomologies and anomalies in supersymmetric theories,''
  Nucl.\ Phys.\ B {\bf 252}, 458 (1985).

\bibitem{BK86}
  I.~L.~Buchbinder and S.~M.~Kuzenko,
  ``Matter superfields in external supergravity: Green functions, effective action and superconformal anomalies,''
  Nucl.\ Phys.\ B {\bf 274}, 653 (1986).
  
\bibitem{K13} 
  S.~M.~Kuzenko,
  ``Super-Weyl anomalies in N=2 supergravity and (non)local effective actions,''
  JHEP {\bf 1310}, 151 (2013)
  [arXiv:1307.7586 [hep-th]].

\end{thebibliography}
\end{document}